\documentclass[reprint,twocolumn,amsmath,amssymb,aps,superscriptaddress,prb,eqsecnum]{revtex4-2}
\usepackage{graphicx}
\usepackage{bm}
\usepackage{epsfig}
\usepackage{amssymb}
\usepackage{amsfonts}
\usepackage{amsmath}
\usepackage{color}
\usepackage{xcolor}
\usepackage{epstopdf}
\usepackage{hyperref}
\usepackage{float}
\usepackage{appendix}
\usepackage{wasysym}
\restylefloat{table}
\usepackage{bibentry}
\usepackage{multirow}
\usepackage[caption=false]{subfig}
\usepackage{braket}
\usepackage{bbm}
\newcommand{\ba}{\begin{eqnarray}}
\usepackage{manfnt}
\newcommand{\ea}{\end{eqnarray}}
\newcommand{\bd}{\begin{displaymath}}
\renewcommand{\v}[1]{{\bf #1}}
\newcommand{\nn}{\nonumber \\}
\newcommand{\md}{~{\rm mod}~}

\newcommand{\rom}[1] {\romannumeral #1}

\DeclareGraphicsExtensions{.pdf,.png,.jpg}

\begin{document}

\title{Rank-2 Toric Code in Two Dimensions}

\author{Yun-Tak Oh}
\thanks{These authors contributed equally to this work.}
\affiliation{Department of Physics, Korea Advanced Institute of Science and Technology, Daejeon 34141, Republic of Korea}

\author{Jintae Kim}
\thanks{These authors contributed equally to this work.}
\affiliation{Department of Physics, Sungkyunkwan University, Suwon 16419, Korea}

\author{Eun-Gook Moon}
\affiliation{Department of Physics, Korea Advanced Institute of Science and Technology, Daejeon 34141, Republic of Korea}
\author{Jung Hoon Han}
\email[Electronic address:$~~$]{hanjemme@gmail.com}
\affiliation{Department of Physics, Sungkyunkwan University, Suwon 16419, Korea}

\date{\today}
\begin{abstract}
We study a two-dimensional spin model obtained by ``Higgsing'' the rank-2 U(1) lattice gauge theory (LGT) with scalar or vector charges on the $L_x \times L_y$ square lattice under the periodic boundary condition (PBC). There are $p$ degrees of freedom per orbital and three orbitals per unit cell in the spin model. The resulting spin model is a stabilizer code consisting of three mutually commuting projectors that are, in turn, obtained by Higgsing the mutually commuting Gauss's law operators and the magnetic field operators in the underlying LGT. The spin model thus obtained is exactly solvable, with the ground state degeneracy (GSD) $D$ given by $\text{log}_p D=2+(1+\delta_{L_x~ \text{mod} ~ p,0})(1+\delta_{L_y~\text{mod}~ p,0})$ when $p$ is a prime number. Two types of dipole excitations, pristine and emergent, are identified. Both the monopoles and the dipoles are free to move, with restrictions on monopoles to hop only by $p$ lattice spacing along with certain directions. The monopole-monopole braiding phase depends on the separation of the $x$ or $y$ coordinates of the initial monopole positions, making it distinct from the ordinary anyon braiding statistics. The monopole-dipole braiding obeys the usual anyonic statistics. Despite the oddity, the monopole-monopole braiding phase can be understood as the Aharonov-Bohm phase of some emergent vector potentials.
\end{abstract}
\maketitle

\section{Introduction}
Fractons have come to embody excitations in lattice models that are immobile on their own~\cite{haah11, vijay15, chamon05, hermele19}. Many variations of the fracton model such as Haah's cubic code and X-cube model have been proposed so far, and new ones continue to show up~\cite{haah11, vijay15}. A defining property common to all fracton models is, besides its restricted mobility of quasiparticles, the sub-extensive GSD, although recently a different kind of fracton model with extensive GSD has been suggested~\cite{kim21}. Field-theoretic interpretations of the fracton physics have been in development for some time~\cite{slagle17, hermele18b, barkeshli18, you20a, chamon21} employing various modifications of the Chern-Simons and/or BF theories. Another line of thinking was initiated by Pretko~\cite{pretko17, pretko172}, who pointed out that the rank-2 U(1) gauge theory developed by Xu and collaborators~\cite{xu06prb, xu16} offers an intuitive understanding of the sub-dimensional motion of quasiparticle excitations in fracton models. 

In ordinary electrodynamics, charges are created as a charge-neutral dipole and subsequently, each charge executes a free motion. Such processes are allowed within the constraints of the rank-1 U(1) gauge theory and the charge conservation (Gauss's) law. A rank-2 U(1) gauge theory~\cite{pretko17,pretko172} typically imposes both charge and dipole conservations, permitting the creation of a dipole-antidipole pair but not the individual dipole. Subsequently, eaach dipole can move freely but not the individual charge that comprise it. By ``Higgsing'' the rank-2 U(1) LGT as was done in Refs.~\onlinecite{hermele18b, barkeshli18}, the constraints become somewhat relaxed and the free motion of the charge can take place as well. 

In this paper, we re-visit the Higgsing procedure and write down an exactly solvable, stabilizer spin Hamiltonian in two dimensions. The idea is, similar to previous works~\cite{hermele18b, barkeshli18} and in particular Ref. \onlinecite{barkeshli18}, to identify some operators in the parent rank-2 U(1) LGT that commute with one another and then elevate them to mutually commuting spin operators through the exponentiation procedure commonly known as ``Higgsing''. In short, the Higgsing procedure converts the rank-2 U(1) LGT, well known to be unstable to the proliferation of instanton in two dimensions~\cite{xu16,pretko17,barkeshli18} into a gapped ${\mathbb Z}_p$ spin model with $p$ chosen to be some prime number. The case of $p=2$ and even-by-even lattice dimensions $L_x \times L_y$ was explored in Ref. \onlinecite{barkeshli18}. The spin model obtained in this manner will be dubbed  `rank-2 toric code (R2TC)' as it is obtained from Higgsing the rank-2 U(1) gauge theory. The well-known toric code is obtained by Higgsing the rank-1 U(1) LGT instead and warrants the classification as the rank-1 toric code (R1TC). 

The nature of elementary quasiparticles in the R2TC model consisting of monopoles and dipoles and their mutual statistics are analyzed in detail. Whereas the charge excitations in the rank-2 LGT are completely immobile due to the constraint of dipole moment conservation, monopole excitations in the R2TC can move in any direction due to the relaxation of the constraint. The quasiparticles of R2TC are not as free as the anyonic quasiparticles of the R1TC, though, in that they are bound to hop only by $p$ lattice spacings in certain directions. Such restriction is however lifted when the two monopoles combine to form a dipole, whose motion is completely free. Most surprisingly, the braiding statistics between a pair of monopoles in our R2TC model is not that of ordinary abelian anyons. Rather, {\it the braiding phase depends on the separation of $x$ or $y$ coordinates of the initial monopole positions}. Such feature is distinct from path-independent anyonic braiding statistics in R1TC. The braiding of a dipole round a monopole, on the other hand, does exhibit the familiar anyonic statistics. The dipole-dipole braiding statistics is bosonic. In addition, we show that the GSD of the R2TC depends on the (mod $p$) linear system sizes $L_x$ and $L_y$.

We begin by reviewing how the Higgsing procedure transforms the rank-1 U(1) LGT into the well-known toric code in Sec. \ref{sec:2}. Afterward, we review the rank-2 U(1) LGT in Sec. \ref{sec:3} and construct the rank-2 toric code by applying the Higgsing procedure in Sec. \ref{sec:4}. Although the Higgsing recipe itself works for both two and three dimensions, detailed analyses are performed for two-dimensional models only due to the difficulty of analyzing three-dimensional models to the same depth. We work out GSD of the R2TC in Sec. \ref{sec:5} and all the related logical operators in Sec. \ref{sec:6}, both of which turn out to depend on the mod $p$ parity of the linear dimensions $L_x,~ L_y$ of the square lattice under the PBC. After identifying all the monopole as well as dipole quasiparticle excitations of the model in Sec. \ref{sec:7}, we work out their mutual braiding statistics in Sec. \ref{sec:8}. In Sec. \ref{sec:9}, we show that the position-dependent phase of the mutual statistics of the monopoles can be described by the integral of the magnetic field. Summary of our work is given in Sec. \ref{sec:10}. 

\section{Toric code from Higgsing the Rank-1 U(1) LGT}
\label{sec:2}

In this section, we review the rank-1 U(1) LGT in two-dimensional square lattice and show how the `Higgsing' procedure transforms it to the well-known R1TC. The rank-1 U(1) LGT is defined in terms of a pair of canonical variables - the compact gauge field $A_i^a$ and the electric field $E_i^a$ ($a=x,~y$) assigned on the links (labeled by $i$) of the square lattice. The two fields obey the canonical commutation relation,
\begin{align}
\left[A_i^a, E_j^b \right] = i \delta_{i,j} \delta_{a,b}.
\label{eq:R1LGT-can-conj}
\end{align}
The compactness comes from the gauge field being an angular variable $A_i^a \in [0,2\pi)$. Consequently, $E_i^a  \sim -i\partial / \partial A_{i}^a$ takes on integer eigenvalues. 

\begin{figure}[h]
\includegraphics[width=0.4\textwidth]{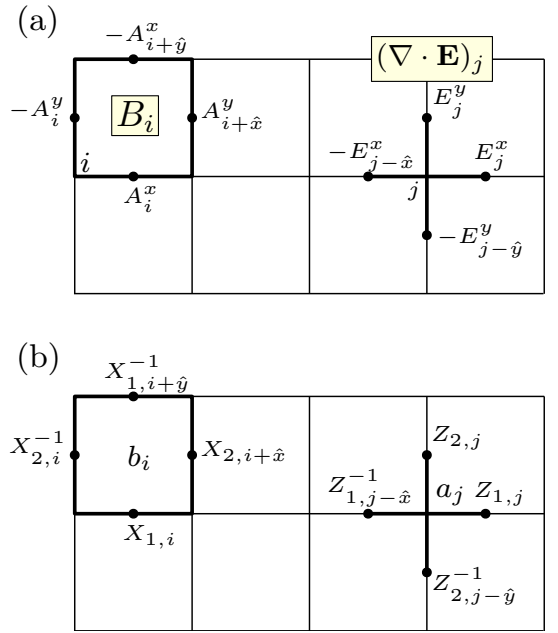}
\caption{
(a) Magnetic field $B_i$ and the Gauss's law $\left({\bm \nabla} \cdot {\bm E}\right)_j$ of rank-1 U(1) LGT. (b) $b_i$ and $a_j$ operators of the rank-1 toric code. Higgsing procedure transforms $B_i$ and $\left({\bm \nabla} \cdot {\bm E}\right)_j$ into $b_i$ and $a_j$ operators, respectively, of the toric code. }
\label{fig:r1lgt}
\end{figure}

On two-dimensional square lattice, the magnetic field $B_i$ of an elementary plaquette in the rank-1 U(1) LGT is defined by
\begin{align}
B_i = A_{i+\hat{x}}^y - A_i^y - A_{i+\hat{y}}^x - A_i^x 
\label{eq:R1LGT-bfield}
\end{align}
as illustrated in Fig. \ref{fig:r1lgt}(a). The local change of the gauge field is implemented as the unitary transformation 
\begin{align}
A_i^a & \rightarrow e^{-i \sum_j f_j ({\bm \nabla} \cdot {\bm E})_j }A_i^a e^{i \sum_j f_j ({\bm \nabla} \cdot {\bm E})_j } \nn 
& = A_i^a + f_{i+\hat{a}}-f_i, \label{eq:local-U1-for-R1}
\end{align}
where $f_i$ is an arbitrary scalar field at the link $i$. The generator for the gauge transformation is
\begin{align}
\left({\bm \nabla} \cdot {\bm E}\right)_i \equiv  E^x_{i}-E^x_{i-\hat{x}} +  E^y_{i}-E^y_{i-\hat{y}},
\label{eq:R1LGT-Gauss}
\end{align}
which is often known as the Gauss's law.

Note that the Gauss's law $\left( {\bm \nabla} \cdot \v {\bm E} \right)_i$ and the magnetic field $B_j$ commute:
\begin{align}
\left[ \left( {\bm \nabla} \cdot {\bm E}\right)_i, B_j \right]=0.
\end{align}
Such commutativity of the magnetic field and the Gauss's law is general in gauge theories, both discrete and continuous, as it comes from the fact that the Gauss's law is the generator of the relevant gauge transformation and the magnetic field is a quantity which, by definition, is invariant under such gauge transformation. A simple recipe emerges for constructing a family of exactly solvable spin models based on the commuting elements of the parent LGTs. The idea is to exploit the commutative nature of the Gauss's laws and the magnetic fields of the parent LGT to construct mutually commuting spin operators, and from there, some exactly solvable models in the form of the stabilizer Hamiltonian~\cite{gottesman96, gottesman97, kitaev03}.

The first step in the construction is to define the local ${\mathbb Z}_p$ Hilbert space consisting of states: $|g\rangle = |0\rangle,~ |1\rangle,~ \cdots,~ |p-1\rangle$. Next, one exponentiates the two fields $(A, E)$ in the rank-1 U(1) LGT as
\begin{align}
    X = e^{ i A } , ~~~ Z = e^{ 2\pi i E /p} . 
    \label{eq:discrete}
\end{align}
The canonical commutation $[A, E] = i$ gives rise to the algebra
\begin{align}
X|g\rangle = |g+1 \rangle, ~ Z|g\rangle  = \omega^g |g\rangle, ~ ZX = \omega XZ,
\label{eq:Z3-spin}
\end{align}
where $\omega = e^{i 2\pi /p}$ and all additions are mod $p$. The basis states $|g\rangle$ are naturally given as eigenstates of $Z$.

The next step is the exponentiation of the Gauss's law and the magnetic field themselves according to the recipe in Eq. (\ref{eq:discrete}):
\begin{align}
a_i &\equiv \exp\left( \frac{2 \pi i}{p}  ({\bm \nabla} \cdot {\bm E})_i \right)  = Z_{1,i} Z_{1,i-\hat{x}}^{-1} Z_{2,i} Z_{2,i-\hat{y}}^{-1}  \nn
b_i  &\equiv \exp( i B_i) = X_{1,i} X_{2,i+\hat{x}} X^{-1}_{1,i+\hat{y}} X^{-1}_{2,i}.
\end{align}
The connection between the gauge theory operators and the spin operators through the exponentiation procedure is illustrated in Fig. \ref{fig:r1lgt}. Note that we are using subscript 1 (2) for the spin operators at the horizontal (vertical) links. The Gauss's law in Eq. (\ref{eq:R1LGT-Gauss}) is the directed sum of the $E$ field operators at the four links emanating from a vertex. When exponentiated, it becomes the $a_i$ operator, a product of $Z$'s and $Z^{-1}$'s on the four links as shown in Fig. \ref{fig:r1lgt}. This is one of the operators one can use to construct the stabilizer. The magnetic field, on the other hand, is the directed sum of the gauge field operators $A$ around an elementary plaquette which, upon exponentiation and applying Eq. (\ref{eq:discrete}), becomes the $b_i$ operator consisting of the product of $X$'s and $X^{-1}$'s. This becomes the second set of stabilizers. The Hamiltonian is then given as the sum of the two types of stabilizers. We refer to the mapping of the Gauss's laws and the magnetic field into corresponding spin operators as the ``Higgsing" of the lattice gauge theory in accordance with the nomenclature in the recent literature~\cite{fradkin79prd,hermele18b, barkeshli18}. Once the Higgsing procedure is complete, one can do away with the parent LGT and focus on the analysis of various interesting properties of the resulting spin model.

The commutativity $\left[ \left( {\bm \nabla} \cdot {\bm E}\right)_i, B_j \right]=0$ of the parent LGT is inherited as that of the spin operators, $[a_i , b_j ] =0$.
Mutually commuting projectors are then constructed,
\begin{equation}
\mathbb{A}_i = \frac{1}{p} \sum_{j =0}^{p-1} (a_i )^j  , ~~ \mathbb{B}_i = \frac{1}{p} \sum_{j =0}^{p-1} (b_i )^j .
\end{equation}
One can prove their projector properties, $\mathbb{A}_i^2 = \mathbb{A}_i$, etc. The Hamiltonian $H = -\sum_i (\mathbb{A}_i + \mathbb{B}_i)$ obtained in this way is none other than the $\mathbb{Z}_p$ toric code. The Higgsing scheme makes transparent the fact that the exact solvability of the toric code originates from the commuting relation of the Gauss's law and the magnetic field in the parent LGT. It is also clear that the scheme will generalize to other LGTs to generate new, exactly solvable spin models.

\begin{figure}[tb]
\includegraphics[width=0.42\textwidth]{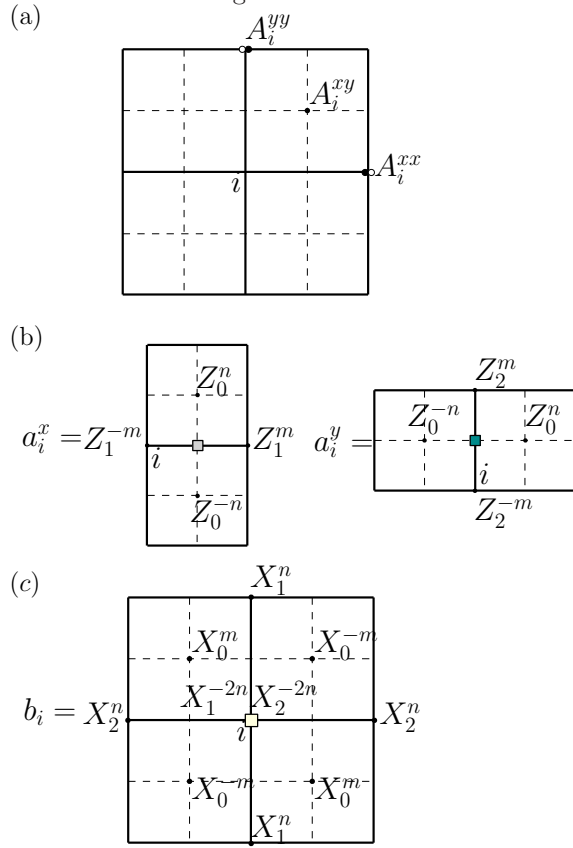}
\caption{(a) Arrangement of the gauge fields for a rank-2 U(1) LGT on square lattice. 
Two circles at lattice site and one circle on the center of plaquette represent three types of gauge fields, respectively.
In particular, the fields labeled by the vertex $i$, $A_i^{xy}$, $A_i^{xx}$ and $A_i^{yy}$, are represented by the black circles. The spin operators (b) $a_{i}^x$, $a_i^y$, and (c) $b_i$ obtained from Higgsing the Gauss's laws and the magnetic field in the rank-2 U(1) LGT with vector charge. Corresponding quasiparticle excitations are shown as filled squares. Operators with subscript 0 are defined at the plaquette, and those with subscripts 1, 2 at the vertices. 
}
\label{fig:2d-lattice-gauge}
\end{figure}

\section{Rank-2 U(1) LGT in two dimensions}
\label{sec:3}
The rank-2 U(1) LGT~\cite{xu06prb} can be defined on a two-dimensional square lattice with compact gauge fields $A_i^{ab}$ $(a,b=x,y)$ assigned on vertices ($A^{xx}_i$ and $A^{yy}_i$) as well as on centers of the plaquettes ($A^{xy}_i =A^{yx}_i$) as depicted in Fig. \ref{fig:2d-lattice-gauge}(a). There are altogether three gauge fields ($A_i^{xx}, A_{i}^{xy}, A_i^{yy}$) per vertex $i$. Note the slightly unconventional assignment of $A_i^{xx}$ and $A_i^{yy}$ on two different vertices as shown in Fig. \ref{fig:2d-lattice-gauge}.

One can write down two kinds of rank-2 U(1) LGTs depending on whether the scalar or the vector charge is assumed~\cite{xu16, pretko17, hermele18b, barkeshli18}. In both cases, the gauge field $A_i^{ab}$ and the electric field $E_i^{ab}$ obey the commutation relation,
\begin{equation} 
[A_i^{xx},E_j^{xx}] = [A_i^{yy},E_j^{yy}] = [A_i^{xy},E_j^{xy}] =  i \delta_{ij}.\label{eq:commutation-algebra} 
\end{equation}
For the rank-2 U(1) LGT with scalar charge $\rho_i$, the magnetic fields and the Gauss's law are given respectively by~\cite{barkeshli18}
\begin{align}
B_i^x &\equiv  m ( A_{i}^{xy} - A_{i-\hat{x}}^{xy} )  - n ( A_{i-\hat{x}+\hat{y}}^{xx} -  A_{i-\hat{x}}^{xx}), \nn
B_i^y & \equiv n ( A_{i+\hat{x}-\hat{y}}^{yy} - A_{i-\hat{y}}^{yy} ) - m ( A_{i}^{xy} - A_{i-\hat{y}}^{xy} ),
\label{eq:r2lgt-scalar-bfield}
\end{align}
and
\begin{widetext}
\begin{align}
({\bf D} E)_i =  m\Bigl(E_{i-2\hat{x}}^{xx} - 2 E_{i-\hat{x}}^{xx} + E_{i}^{xx} +E_{i-2\hat{y}}^{yy}- 2 E_{i-\hat{y}}^{yy} + E_{i}^{yy} \Bigl) + n\left(E_{i-\hat{x}-\hat{y}}^{xy}  - E_{i-\hat{y}}^{xy} - E_{i-\hat{x}}^{xy} + E_{i}^{xy} \right) \equiv \rho_i \label{eq:gauss}
\end{align}
\end{widetext}
with $(m,n)$ integers. The ``tensor divergence'' $({\bf D} E )_i$ generalizes the vector divergence $({\bm \nabla} \cdot {\v E})_i$ of the rank-1 U(1) gauge theory. 

Vital to the later construction is their commutativity $\left[ ( {\bf D} E )_i , B_j^x \right] = \left[ ( {\bf D} E )_i , B_j^y \right] = 0$, which can be checked explicitly from their respective definitions in Eqs. (\ref{eq:r2lgt-scalar-bfield}) and (\ref{eq:gauss}). The local gauge transformation rule for the rank-2 U(1) LGT with scalar charge is
\begin{align}
A_i^{ab} & \rightarrow e^{-i \sum_j f_j ({\bf D} E)_j }A_i^{ab} e^{i \sum_j f_j ({\bf D} E)_j } \nn 
& = A_i^{aa} -m(f_{i+2\hat{a}}-2f_{i+\hat{a}}+f_i) &(a&=b)\nn
& = A_i^{ab} -n(f_{i+\hat{a}+\hat{b}}-f_{i+\hat{a}}-f_{i+\hat{b}}+f_i) &(a&\neq b) . \end{align}
One can readily check that the magnetic fields ($B_i^x , B_i^y)$ in Eq. (\ref{eq:r2lgt-scalar-bfield}) are invariant under this transformation. 

For the rank-2 U(1) LGT with vector charge, there is one component of the magnetic field 
\begin{widetext}
\begin{align} 
 B_i \equiv  n (A_{i-\hat{x}-\hat{y}}^{xx} - 2 A_{i-\hat{x}}^{xx} + A_{i-\hat{x}+\hat{y}}^{xx} +A_{i-\hat{x}-\hat{y}}^{yy} - 2 A_{i-\hat{y}}^{yy} + A_{i+\hat{x}-\hat{y}}^{yy} )
- m (A_i^{xy}  - A_{i-\hat{y}}^{xy} - A_{i-\hat{x}}^{xy} + A_{i-\hat{x}-\hat{y}}^{xy} ),
\label{eq:b-field}
\end{align}
\end{widetext}
and the two Gauss's laws
\begin{align}
( {\bf D} E )_i^{x} & =  m(E_{i+x}^{xx}-E_{i}^{xx}) + n(E_{i+y}^{xy}-E_{i}^{xy}) \equiv \rho_i^x, \nn
( {\bf D} E )_i^{y} & =  m (E_{i+y}^{yy}-E_{i}^{yy}) + n ( E_{i+x}^{yx}- E_{i}^{yx}) \equiv \rho_i^y ,\label{eq:vector-gauss-laws}
\end{align}
with commutation $\left[ ( {\bf D} E )_i^{x}, B_j \right] = \left[ ( {\bf D} E )_i^{y}, B_j \right] = 0$. The local U(1) transformation in the vector charge theory reads
\begin{align}
A_i^{ab} & \rightarrow e^{-i \sum_{j,c} [f^c_j ({\bf D} E)^c_j ] }A_i^{ab} e^{i \sum_{j, c} [f^c_j ({\bf D} E)^c_j] } \nn 
& = A_i^{aa} -m(f^a_{i+\hat{a}}-f^a_i) &&(a=b)\nn
& = A_i^{ab} -n(f^b_{i+\hat{a}}-f^b_i+f^a_{i+\hat{b}}-f^a_i) &&(a\neq b),
\end{align}
which leaves $B_i$ invariant. 

These two types of rank-2 U(1) LGTs have been discussed in the past~\cite{pretko17, pretko172, barkeshli18}. In particular, Gauss's laws of the rank-2 U(1) LGT enforce the restricted mobility of the monopole excitations associated with $\rho_i$, $\rho_i^x$, and $\rho_i^y$, qualifying them as fractons $(\rho_i)$ and lineons ($\rho_i^x$ and $\rho_i^y$)~\cite{vijay16, xu16}. The dipolar quasiparticles formed as a pair of oppositely charged fractons or lineons, on the other hand, are not restricted by Gauss's law and can move freely.

\begin{figure*}[tb]
\includegraphics[width=0.99\textwidth]{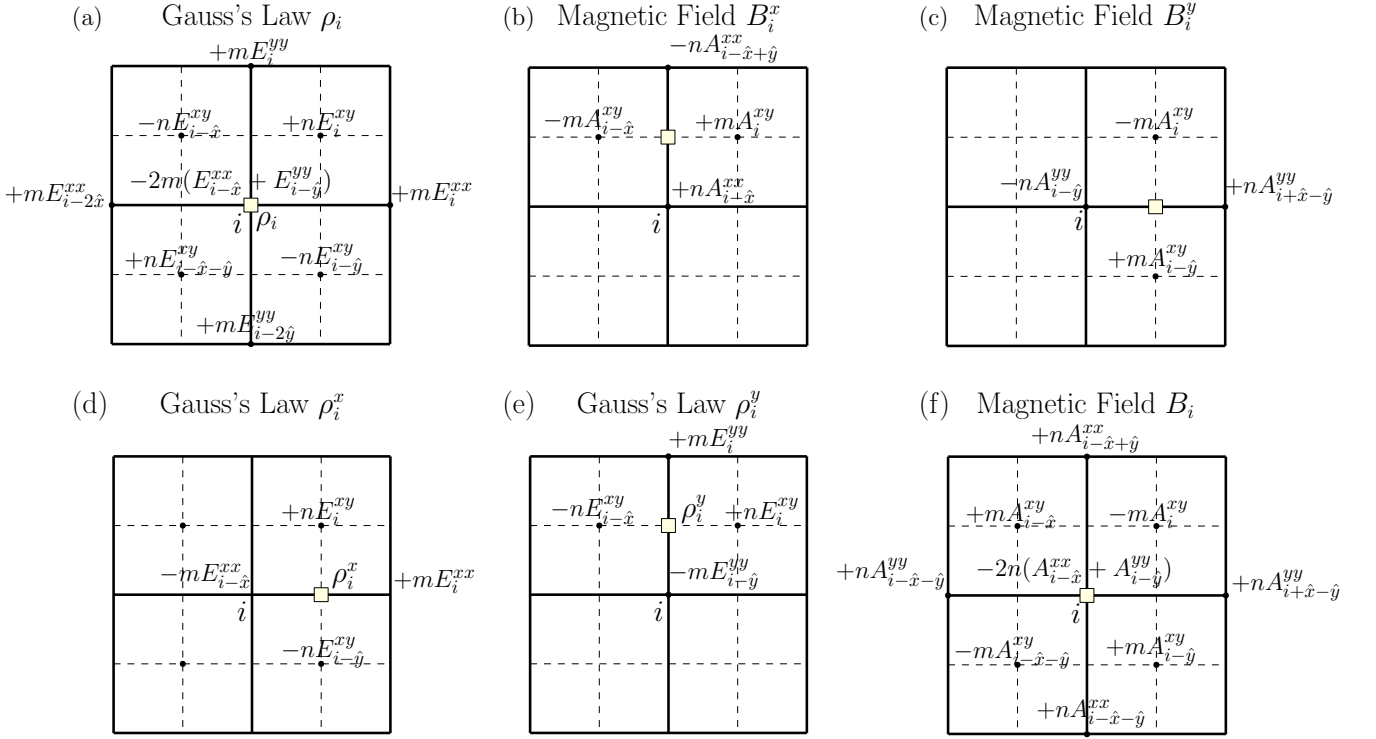}
\caption{Gauss's laws and magnetic fields in the rank-2 U(1) LGT with (first row) scalar charge and (second row) vector charge. (a) Gauss's law. The scalar charge $\rho_i$ is defined at the vertex $i$. (b) and (c) Two magnetic fields $B_i^x$ and $B_i^y$. (d) and (e) Two Gauss's laws associated with the two vector charges $(\rho^x_i , \rho^y_i )$. The vector charges $\rho_i^x$ and $\rho_i^y$ are defined on the links. (f) Magnetic field $B_i$ of the rank-2 U(1) LGT with vector charge.}
\label{fig:SV-gauss-law}
\end{figure*}

In fact, the vector and the scalar charge theories are dual to each other in two dimensions and, as a result, only one theory (e.g. vector charge theory) needs to be analyzed in detail. In Fig. \ref{fig:SV-gauss-law}, Gauss's laws and the magnetic fields of scalar and vector charge rank-2 U(1) LGT models are depicted. By comparing them one can easily recognize a definite relation between the Gauss's laws and the magnetic fields in one theory with those in the other. Specifically, the mapping
\begin{align}
( m E_i^{xx} , n A_i^{xx}  )_s &\leftrightarrow  ( n A_{i+\hat{x}-\hat{y}}^{yy} , - m E_{i-\hat{y}}^{yy} )_v  \nn
( m E_i^{yy} , n A_i^{yy} )_s &\leftrightarrow ( n A_{i-\hat{x}+\hat{y}}^{xx} , - m E_{i-\hat{x}+\hat{y}}^{xx})_v \nn
( n E_i^{xy} , m A_i^{xy} )_s &\leftrightarrow ( -m A_{i}^{xy} , n E_{i}^{xy} )_v , \label{eq:dual-mapping}
\end{align}
between the operators in the scalar (subscript $s$) and the vector (subscript $v$) charge theories converts the Gauss's law for $\rho_i$ and the magnetic field $(B^x_i , B^y_i)$ in the scalar theory to the magnetic field $B_{i}$ and the charges $(\rho_{i}^y , - \rho_{i}^x)$ in the vector theory, respectively. The duality applies only in two dimensions.

\section{Rank-2 Toric Code}
\label{sec:4}
We exploit the Higgsing procedure detailed in Sec. II to construct the stabilizer spin Hamiltonian out of the Gauss's law and the magnetic field of the parent rank-2 U(1) LGT with vector charge given in Eqs. (\ref{eq:b-field}) and (\ref{eq:vector-gauss-laws}). The final result is the rank-2 analogue of the toric code, which we call the rank-2 toric code, or R2TC for short. 

Following Eq. (\ref{eq:discrete}),  we convert the canonical variables $A_i^{ab}$ and $E_i^{ab}$ into spin operators as
\begin{align}
X_{0,i} = e^{i A_i^{xy}}, & ~~ Z_{0,i} = e^{2\pi i E_i^{xy}/p } \nn 
X_{1,i} = e^{i A_i^{xx}}, & ~~ Z_{1,i} = e^{ 2\pi i E_i^{xx}/p}, \nn 
X_{2,i} = e^{i A_i^{yy}}, & ~~ Z_{2,i} = e^{ 2\pi i E_i^{yy}/p}. 
\end{align}
The two indices $1$ and $2$ represent the two orbitals assigned on the vertices and the index $0$ the orbital on the plaquette center. The second step is the exponentiation of the Gauss's law and the magnetic field in Eqs. (\ref{eq:b-field}) and (\ref{eq:vector-gauss-laws}) according to the recipe in Eq. (\ref{eq:discrete}):
\begin{align}
    a_i^x &\equiv \exp \left( \frac{2 \pi i}{p}  \rho_i^x \right) =Z_{1,i}^{-m} Z_{1,i+\hat{x}}^{m} Z_{0,i}^n Z_{0,i-\hat{y}}^{-n},\nn
    a_i^y &\equiv  \exp\left( \frac{2 \pi i}{p}  \rho_i^y \right)=Z_{2,i}^{-m} Z_{2,i+\hat{x}}^{m} Z_{0,i}^n Z_{0,i-\hat{x}}^{-n},\nn
    b_i  &\equiv \exp( i B_{i})\nn
&= X_{2,i-\hat{x}}^n X_{2,i}^{-2n} X_{2,i+\hat{x}}^n 
X_{1,i-\hat{y}}^n X_{1,i}^{-2n} X_{1,i+\hat{y}}^n \nn
&~~~\times X_{0,i}^{m} X_{0,i-\hat{x}}^{-m} X_{0,i-\hat{y}}^{-m} X_{0,i-\hat{x}-\hat{y}}^{m}. 
    \label{eq:vector_Zn-operators}
\end{align}
They are illustrated in Fig. \ref{fig:2d-lattice-gauge}(b). The commutativity among $(\rho^x_i, \rho^y_i, B_i)$ is inherited now as that of spin operators $(a_i^x, a_i^y , b_i )$. This property allows us in turn to construct a stabilizer Hamiltonian as
\begin{align}
    \mathbb{A}_i^x & = \frac{1}{p} \sum_{j =0}^{p-1} (a_i^x )^j ,    &\mathbb{A}_i^y & = \frac{1}{p} \sum_{j =0}^{p-1} (a_i^y )^j , \nn
    \mathbb{B}_{i} & = \frac{1}{p} \sum_{j =0}^{p-1} (b_i )^j ,    &H & = -\sum_i (\mathbb{A}^x_i + \mathbb{A}^y_i + \mathbb{B}_i ) . 
    \label{eq:2d-lattice-model}
\end{align}
This completes the construction of the R2TC - the exactly solvable stabilizer model from the underlying rank-2 U(1) LGT. The ground state(s) is characterized by $\mathbb{A}_i^x |{\rm GS}\rangle = \mathbb{A}_i^y |{\rm GS} \rangle = \mathbb{B}_i |{\rm GS} \rangle = |{\rm GS} \rangle$. 

\section{Ground State Degeneracy}
\label{sec:5}
Calculation of the GSD for exactly solvable spin models is based on the general formula~\cite{gottesman96, shor97, kim21},
\begin{align}
    \log_p D = N_o - N_s = N_{lo},
    \label{eq:GSD_formula}
\end{align}
where $D$, $p$, $N_o$, $N_s$, and $N_{lo}$ represent the GSD, the local Hilbert space dimension, the number of orbitals, the number of independent stabilizers, and the number of independent logical operators, respectively. There are three orbital degrees of freedom, two at the site and one at the dual lattice site, for $N_{o} = 3 L_x L_y$ on a $L_x \times L_y$ square lattice with PBC. 

We consider the case of $p$ being a prime number in this work due to the extra complication caused by non-prime integer $p$. For example, if $p~{\rm mod}~m=0$ in the $(m,n)$ vector charge theory, we have the identity
\begin{align}
\left( a_i^x (a_i^y )^{-1} ( a_{i-\hat{x}}^x )^{-1} a_{i-\hat{y}}^y \right)^{p/m} = 1.
\end{align}
This identity, which holds for every site $i$ of the lattice, creates constraints that will have to be taken into account when we later try to count the number of independent stabilizers. Furthermore, operators such as  $Z_{0,i}^{p/m}$, $X_{1,i}^{p/m}$, and $X_{2,i}^{p/m}$ commute with the Hamiltonian and act as local symmetry generators, which vastly affect the ground state structure. For these reasons, $p$ is a prime number in the remainder of the paper. 

The integers $(m,n)$ are, of course, only meaningful mod $p$, since the $p$-th power of either $X$ or $Z$ operator is an identity. This is also the meaning of ``Higgsing", by which the strict charge conservation of the underlying rank-2 U(1) LGT relaxes to the mod $p$ conservation. So far, the pair of integers $(m,n)$ were kept general. Keeping them was helpful in identifying the duality relation between scalar and vector charge theories in Sec. \ref{sec:3}. In the following, however, we restrict ourselves to the case $(m,n)=(1,1)$ in enumerating the GSD and for other analyses of the model. We believe the essence of the subsequent arguments and analyses remain insensitive to the choice of $(m,n)$. 

The GSD of the R2TC calculated for the $L_x \times L_y$ square lattice under the PBC turns out to depend on whether $L_x$ or $L_y$ is a multiple of $p$. The following products of operators prove useful in counting the GSD:
\begin{align}
H_{j_y} (O) &= \prod_{j_x = 1}^{L_x} O_{j_x \hat{x} + j_y \hat{y}},   \nn 
V_{j_x}(O) &= \prod_{j_y = 1}^{L_y} O_{j_x \hat{x}+j_y\hat{y}},\nn
M(O) &= \prod_{j_x=1}^{L_x} \prod_{j_y=1}^{L_y} O_{j_x\hat{x}+j_y \hat{y}}.
\end{align}
Here, $O$ is an operator in the R2TC model. As one can see, $H_{j_y}(O)$ is a product of $O$'s along the $j_y$-th row, and $V_{j_x}(O)$ the product of $O$'s along the $j_x$-th column. $M(O)$ is referred to as a membrane operation since it is the product of $O$'s over the entire lattice. 

In addition to the horizontal and the vertical operators defined above, some diagonal operators can be defined by
\begin{align}
D_i^{xy}(O) & \equiv \prod^{L_l}_{j=1} O_{i+j(\hat{x} +\hat{y})} , \nn 
D_i^{x\bar{y}}(O) & \equiv \prod^{L_l}_{j=1} O_{i+j(\hat{x} - \hat{y})}.
\end{align}
They are the products of $O$'s along the $(1,1)$ or $(1,-1)$ direction starting from the site $i$. The site coordinates are labeled modulo $( L_x, L_y)$ under the PBC. The product over $j$ in the above diagonal operators spans $1 \le j \le L_l$ where $L_l = {\rm lcm} (L_x , L_y)$ is the least common multiple of $L_x$ and $L_y$.  Note that $L_l$ is the smallest number for which $O_{i+j(\hat{x}+\hat{y})}=O_{i}$. There are $L_g$ distinct diagonal operators, where $L_g = {\rm gcd} (L_x, L_y)$ stands for the greatest common divisor. Specifically, we have $D_{i}^{xy}(O),~ D_{i+\hat{y}}^{xy}(O),\cdots,~ D_{i + ( L_g -1) \hat{y}}^{xy} (O)$ in the (1,1) direction. In total, due to $L_xL_y=L_gL_l$, every site appears exactly once in the diagonal operator of a given direction. 

At first there seem to be $3L_x L_y$ stabilizers, $a_{i}^x$, $a_i^y$, and $b_i$, for all $i$, on a $L_x \times L_y$ under the PBC. Some constraints readily emerge amongst them:
\begin{align}
    \prod_{i} a_{i}^x = 1, ~~~ \prod_{i} a_{i}^y = 1, ~~~ \prod_{i} b_{i} = 1 . \label{eq:v-identity-1}
\end{align}
The product $\prod_i$ runs over all the sites, with the identity remaining valid regardless of the lattice size or the degree of freedom $p$, suggesting that the number of independent stabilizers would be $N_s = 3L_x L_y - 3$. Interestingly, several more identities arise among the stabilizers if $L_x$ or $L_y$ is a multiple of the local Hilbert space dimension $p$. 

To figure out what the extra identities are, first consider
\begin{align}
H_{j_y}  (a^x)&  = \prod_{j_x=1}^{L_x} Z_{0,j_x\hat{x} + j_y \hat{y}} Z_{0,j_x\hat{x} + (j_y-1) \hat{y}}^{-1},\nn
V_{j_x}  (a^y)& = \prod_{j_y=1}^{L_y} Z_{0,j_x\hat{x} + j_y\hat{y}} Z_{0,(j_x-1)\hat{x} + j_y\hat{y}}^{-1}.
\end{align}
It is straightforward to show
\begin{align}
&H_{j_y}(a^x) [ H_{j_y-1}(a^x) ]^2\nn
&=\prod_{j_x=1}^{L_x} Z_{0,j_x\hat{x} + j_y \hat{y}} Z_{0,j_x\hat{x} + (j_y-1) \hat{y}} Z_{0,j_x\hat{x} + (j_y-2) \hat{y}}^{-2} , \nonumber 
\end{align}
and in general, 
\begin{align}
\prod_{j_y=1}^{p} \left[H_{-j_y}(a^x)\right]^{j_y} = \prod_{j_y=1}^{p} \prod_{j_x=1}^{L_x } Z_{0, j_x \hat{x} -j_y\hat{y}}
\end{align}
because the last factor $Z_0^p = 1$ vanishes from the product. In the same way we can show
\begin{align}
\prod_{j_x=1}^p \left[V_{-j_x}(a^y)\right]^{j_x} =  \prod_{j_x=1}^{p} \prod_{j_y=1}^{L_y} Z_{0,-j_x \hat{x} +j_y \hat{y} }.
\end{align}
When $L_x$ and $L_y$ are both multiples of $p$, we can extend both of these products to cover the entire lattice and get 
\begin{align}
\prod_{j_y=1}^{L_y} \left[H_{-j_y}(a^x)\right]^{j_y}  &=   \prod_{j_x=1}^{L_x} \left[V_{-j_x}(a^y)\right]^{j_x} 
\label{eq:v-321-identity-2}
\end{align}
This is the first of the extra identities as it relates the product of $a^x$ stabilizers with those of the $a^y$ stabilizers. This identity applies only when both $L_x$ and $L_y$ are multiples of $p$. 

Additional extra identities arise from considering a product of $b_i$ stabilizers:
\begin{align}
\prod_{j_y=1}^p \left[H_{j_y} (b)\right]^{j_y} = \prod_{j_x = 1}^{L_x}
X_{1,j_x\hat{x}} X_{1,j_x\hat{x} + p\hat{y}} ^{-1} . 
\nonumber \end{align}
If $L_y$ is a multiple of $p$, extending the product over $j_y$ over the entire lattice gives the second identity:
\begin{equation}
    \prod_{j_y=1}^{L_y} \left[H_{j_y} (b)\right]^{j_y} = 1. 
    \label{eq:v-321-identity-3}
\end{equation}
This is the second extra identity, which holds only for $L_y \md p = 0$. Arguing along an analogous line gives 
\begin{equation}
    \prod_{j_x=1}^{L_x} \left[V_{j_x} (b)\right]^{j_x} = 1 ,
    \label{eq:v-321-identity-4}
\end{equation}
when $L_x$ is a multiple of $p$. All told, the number of independent stabilizers is $3L_xL_y-2-\left(1+\delta_{0,L_x \mod p}\right)$ $\left(1+\delta_{0,L_y {\rm mod}~p} \right)$, leading to the GSD
\begin{align}
\log_p D &= 2 + \left(1+\delta_{0,L_x~{\rm mod}~p} \right)\left(1+\delta_{0,L_y~{\rm mod}~p} \right) \label{eq:GSD_V_321}
\end{align}
according to the general formula, Eq. (\ref{eq:GSD_formula}). The $\log_p D$ varies from 3 to 6, depending on the mod $p$ parity of $L_x$ and $L_y$. The $2^6$ GSD for $p=2$ in Ref. \onlinecite{barkeshli18} misses the subtle variation taking place with the system size.

\section{Logical operators}
\label{sec:6}

In the previous section, we successfully counted the GSD by utilizing Eq. (\ref{eq:GSD_formula}) and working out the number of independent stabilizers $N_s$ and hence $N_o - N_s$ explicitly. In this section, we identify the logical operators and count their numbers, $N_{lo}$. We show, in accordance with the GSD obtained in the previous section, the number of logical operators generating the different ground states also varies with respect to the system sizes. The logical operators can be constructed either as a product of $Z$'s or of $X$'s. The construction of the two types of logical operators is treated separately. 

\subsection{Logical $Z$ Operators}
The number of independent logical operators made out of the $Z$ operators is 
\begin{align}
    N_{lo} = 2 + (1+\delta_{0,L_x~{\rm mod}~p})(1+\delta_{0,L_y~{\rm mod}~p}),\nonumber
\end{align}
in accordance with Eq. (\ref{eq:GSD_V_321}). They are summed up in Table \ref{tab:Z-logical-operators}. 

\begin{table}[h]
\begin{tabular}{ |l|l|l| }
\hline
& & \\
\multicolumn{1}{|c}{Cases} &\multicolumn{1}{|c}{$L_x~{\rm mod}~p=0$} & \multicolumn{1}{|c|}{$L_x~{\rm mod}~p\neq 0$}\\
& &\\
\hline 
\multicolumn{1}{|c|}{\multirow{3}{*}{$L_y~{\rm mod}~p=0$}} & $H_{j_y}(Z_2),~V_{j_x}(Z_1),$ & 
\multirow{2}{*}{$H_{j_y}(Z_2),~V_{j_x}(Z_1),$}\\
& $H_{j_y}(Z_0),~ V_{j_x}(Z_0),$ & \multirow{2}{*}{$H_{j_y}(Z_0),~\mathbb{V}_{j_x}(Z_1)$}\\
&$\mathbb{H}_{j_y}(Z_2),~\mathbb{V}_{j_x}(Z_1)$ &\\
\hline
\multicolumn{1}{|c}{\multirow{2}{*}{$L_y~{\rm mod}~p\neq 0$}} &  
\multicolumn{1}{|c|}{\multirow{1}{*}{$H_{j_y}(Z_2),~~V_{j_x}(Z_1),$}} &  $H_{j_y}(Z_2),~~V_{j_x}(Z_1),$\\
& \multicolumn{1}{c|}{\multirow{1}{*}{$V_{j_x}(Z_0),~~\mathbb{H}_{j_y}(Z_2)$}} & $M(Z_0)$\\ 
\hline
\end{tabular}
\caption{Independent logical $Z$ operators.}
\label{tab:Z-logical-operators}
\end{table}

Altogether one finds seven types of logical operators in Table \ref{tab:Z-logical-operators}: $H_{j_y}(Z_2)$, $V_{j_x}(Z_1)$, $H_{j_y}(Z_0)$, $V_{j_x}(Z_0)$, $M(Z_0)$, $\mathbb{H}_{j_y}(Z_2)$ and $\mathbb{V}_{j_x}(Z_1)$. All of them consist of a product of $Z$'s and commute with the Hamiltonian. They will be collectively referred to as $Z$-logical operators.

Firstly, we consider the two types of logical operators: $H_{j_y}(Z_2)$ and $V_{j_x}(Z_1)$. These two operators are well defined regardless of the linear system size. The following identities place stringent restrictions on the independence of these operators:  
\begin{align}
    H_{j_y}(a^y) &= \left[H_{j_y-1}(Z_2)\right]^{-1} H_{j_y}(Z_2), \nn
    V_{j_x}(a^x) &= \left[V_{j_x-1}(Z_1)\right]^{-1} V_{j_x}(Z_1).
\end{align}
Since the ground states satisfy $a_i^x |{\rm GS} \rangle = a_i^y |{\rm GS} \rangle = |{\rm GS} \rangle$, the action of $H_{j_y}(a_y)$ or $V_{j_x}(a_x)$ on $|{\rm GS} \rangle$ is an identity, implying $H_{j_y-1}(Z_2) |{\rm GS} \rangle = H_{j_y}(Z_2)|{\rm GS} \rangle$. As a result there is really only one independent logical operator among the $H_{j_y}(Z_2)$'s. Similarly, only one independent operator exists among $V_{j_x } (Z_1 )$'s. We conclude that there are at least two logical operators $V_{j_x} (Z_1 )$ and $H_{j_y} (Z_2)$ regardless of the linear dimensions of the lattice. 

The next type of logical operators consists of a product of $Z_0$'s: $H_{j_y}(Z_0)$, $V_{j_x}(Z_0)$, and $M(Z_0)$. They turn out to have a rather complicated dependence on one another. For starters, $M(Z_0)$ can only be defined when both linear dimensions are incommensurate with $p$: $L_x~{\rm mod}~p \neq 0$ and $L_y~{\rm mod}~p \neq 0$. When one of these conditions fails, for instance $L_y~{\rm mod}~p = 0$, the identity $\prod_{j_y = 1}^{L_y}\left[H_{-j_y}(a^x)\right]^{j_y}  = M(Z_0)$ holds and one gets $M(Z_0 ) |{\rm GS} \rangle = |{\rm GS} \rangle$. Similarly, when $L_x~{\rm mod}~p = 0$, the identity $\prod_{j_x = 1}^{L_x} \left[V_{-j_x}(a^y)\right]^{j_x}  = M(Z_0)$ shows $M(Z_0)$ is trivial. In all, $M(Z_0)$ is a meaningful logical operator only if neither of $L_x ,~ L_y$ is commensurate with $p$. 

For counting the number of independent logical operators amongst $H_{j_y}(Z_0)$ and $V_{j_x}(Z_0)$, we note the identities
\begin{align}
    H_{j_y}(a^x) & = \left[H_{j_y -1}(Z_0)\right]^{-1} H_{j_y}(Z_0),\nn
    V_{j_x}(a^y) & = \left[V_{j_x-1}(Z_0)\right]^{-1} V_{j_x}(Z_0).
\end{align}
Accordingly, there appear two independent logical operators, one from $H_{j_y}(Z_0)$ and one from $V_{j_x}(Z_0)$. However, some additional identities arise among them when $L_y~{\rm mod}~p \neq 0$ or $L_x\md p \neq 0$. Take $L_y \md p\neq0$, for instance, and we can always find $l$ that satisfies $lp \md L_y = 1$ since $p$ is a prime number. For the smallest such $l$, we have
\begin{align}
    \prod_{k=1}^{lp} \left[H_{k + j_y}(a^x)\right]^{-k} = H_{j_y}(Z_0)~ [ M(Z_0) ]^{(lp-1)/L_y}, \label{eq:Z0-identity1}
\end{align}
where the site coordinates are labeled modulo $L_y$ under the PBC. Such relation implies that $H_{j_y}(Z_0) |{\rm GS} \rangle = [M(Z_0 )]^{(1-lp)/L_y} |{\rm GS} \rangle$, where  $(lp-1)/L_y$ is an integer smaller than $p$ by its definition. This implies none of the $H_{j_y} (Z_0)$ logical operators are independent for $L_y \md p \neq 0$ since they can be always constructed in terms of $a_i^x$, $a_i^y$, and $M(Z_0)$. We conclude that $H_{j_y} (Z_0)$ is an independent operator only if $L_y \md p = 0$.
Following a similar procedure, we can show that there is one independent logical operator $V_{j_x}(Z_0)$ only if $L_x \md p = 0$, and none otherwise. In all, the number of independent logical operators amongst $M(Z_0)$, $H_{j_y}(Z_0)$ and $V_{j_X}(Z_0)$ is $1+\delta_{L_x~{\rm mod}~p , 0}\delta_{L_y~{\rm mod}~p , 0}$.

Finally, we have two additional types of logical operators given by
\begin{align}
\mathbb{H}_{j_y}(Z_2) &\equiv \prod_{j_x=1}^{L_x} \left(Z_{2,j_x\hat{x} +j_y\hat{y}}\right)^{j_x}, \nn
\mathbb{V}_{j_x}(Z_1) &\equiv \prod_{j_y=1}^{L_y} \left(Z_{1,j_x\hat{x} +j_y\hat{y}}\right)^{j_y}.
\label{eq:big-H-big-Z}
\end{align}
Note that we have the exponents $j_x$ and $j_y$ on the right-hand sides of Eq. (\ref{eq:big-H-big-Z}), which makes $\mathbb{H}_{j_y}(Z_2)$ and $\mathbb{V}_{j_x}(Z_1)$  distinct from $H_{j_y}(Z_2)$ and $V_{j_x}(Z_1)$ defined previously. Since $(Z_2)^p = 1$, we can decompose $\mathbb{H}_{j_y}(Z_2)$ into blocks, each block consisting of multiplication of $Z_2$ with ascending exponents from $0$ to $p-1$. Each block consisting of $Z_2$'s with ascending exponents corresponds to what we will define as $\mathbb{Z}_{x,i}$ later in Eq. (\ref{eq:Zx-operation2}), which creates a pristine dipole, to be defined in Sec. \ref{sec:q-excitation}. When $L_x \md p = 0$,  $\mathbb{H}_{j_y}(Z_2)$ can be fully decomposed into blocks of $\mathbb{Z}_{x,i}$ and therefore, each pristine dipole created by $\mathbb{Z}_{x,i}$ is annihilated by the subsequent operation of $\mathbb{Z}_{x,i'}$, leaving no residual monopoles. Hence the operator $\mathbb{H}_{j_y}(Z_2)$ commutes with the R2TC Hamiltonian and can become a logical operator, only if $L_x \md p = 0$. Similarly, $\mathbb{V}_{j_x}(Z_1)$ can be a logical operator only when $L_y~ {\rm mod}~p = 0$. 

The following identities need to be invoked while counting the number of independent logical operators amongst $\mathbb{H}_{j_y}(Z_2)$'s and $\mathbb{V}_{j_x}(Z_1)$'s: 
\begin{align}
\prod_{j_x = 1}^{L_x} \left( a^y_{j_x\hat{x} + j_y\hat{y}}\right)^{j_x} & = \left[\mathbb{H}_{j_y}(Z_2)\right]^{-1} \left[H_{j_y}(Z_0)\right]^{-1} \mathbb{H}_{j_y+1}(Z_2), \nn
\prod_{j_y = 1}^{L_y} \left( a^x_{j_x\hat{x} +j_y \hat{y}}\right)^{j_y} & = \left[\mathbb{V}_{j_x}(Z_1)\right]^{-1} \left[V_{j_x}(Z_0)\right]^{-1} \mathbb{V}_{j_x+1}(Z_1).
\label{eq:math-h-v-z-ident}
\end{align}
The first equality holds only if $L_x \md p =0$, in which case $\mathbb{H}_{j_y}(Z_2)$ becomes a valid logical operator. The l.h.s. of the first equality above becomes an identity when acting on the ground state, implying that the action of the logical operator $\mathbb{H}_{j_y+1}(Z_2)$ is equivalent to the combined action of $\mathbb{H}_{j_y}(Z_2)$ in the neighboring row and $H_{j_y}(Z_0)$. Recall that $H_{j_y}(Z_0)$ was already taken into consideration earlier as a logical operator. As a result, we conclude that there is only one independent logical operator among the $\mathbb{H}_{j_y}(Z_2)$'s. Whether $H_{j_y}(Z_0)$ is an independent logical operator or not (depending on $L_y \md p$) does not affect the conclusion. Similarly, one can show that we have one independent operator amongst $\mathbb{V}_{j_x}(Z_1)$'s only when $L_y \md p=0$, and none otherwise. Therefore, the number of independent operators amongst the $\mathbb{H}_{j_y}(Z_2)$'s and $\mathbb{V}_{j_x}(Z_1)$'s become  $\delta_{0,L_x~{\rm mod}~p} + \delta_{0,L_y~{\rm mod}~p}$. Summing up all the arguments thus far gives the correct number for $N_{lo}$.

All of the logical operators are defined on a non-contractible loop except $M(Z_0 )$ in Table \ref{tab:Z-logical-operators}. However, in cases where $M(Z_0)$ does serve as a logical operator it becomes equivalent to $H_{j_y}(Z_0)$ and $V_{j_x}(Z_0)$, both of which are the logical operators defined on non-contractible loops.

\subsection{Logical $X$ Operators}
Similar to the discussion of the logical $Z$ operators in the previous subsection, we begin by listing their explicit forms in Table \ref{tab:X-logical-operators} below. Here one finds five types of logical $X$ operators: $H_{j_y}(X_1)$, $V_{j_x}(X_2)$, $M(X_0)$, $\mathbb{D}_i^{xy}$ and $\mathbb{D}_i^{x\bar{y}}$. 

\begin{table}[h]
\begin{tabular}{ |l|c|l| }
\hline
& & \\
\multicolumn{1}{|c}{Cases} &\multicolumn{1}{|c}{$L_x~{\rm mod}~p=0$} & \multicolumn{1}{|c|}{$L_x~{\rm mod}~p\neq 0$}\\
& &\\
\hline 
\multicolumn{1}{|c|}{\multirow{3}{*}{$L_y~{\rm mod}~p=0$}} &  $H_{j_y}(X_1),~V_{j_x}(X_2),$ & \multirow{2}{*}{$H_{j_y}(X_1),~V_{j_x}(X_2),$}\\
& $H_{j_y+1}(X_1),~ V_{j_x+1}(X_2),$ & \multirow{2}{*}{$H_{j_y+1}(X_1),~M(X_0)$}\\
&$\mathbb{D}_{i}^{xy},~\mathbb{D}_{i}^{x\bar{y}}$ &\\
\hline
\multicolumn{1}{|c}{\multirow{2}{*}{$L_y~{\rm mod}~p\neq 0$}} &  \multicolumn{1}{|c|}{$H_{j_y}(X_1),~~V_{j_x}(X_2),$} &  $H_{j_y}(X_1),~~V_{j_x}(X_2),$\\
& \multicolumn{1}{c|}{$V_{j_x+1}(X_2),~M(X_0)$} & $M(X_0)$ \\
\hline
\end{tabular}
\caption{Independent logical $X$ operators.}
\label{tab:X-logical-operators}
\end{table}

Firstly we consider $M(X_0)$, which is a logical operator except for both $L_x$ and $L_y$ are multiple of $p$. This is in contrast to $M(Z_0)$, which becomes a logical operator only when both $L_x$ and $L_y$ are incommensurate with $p$. First consider the identity
\begin{align}
\prod_{j_x, j_y =1}^p  [b_{(1+j_x)\hat{x} + (1+j_y)\hat{y}}]^{j_x j_y} = \prod_{j_x, j_y =1}^p X_{0,j_x\hat{x} + j_y\hat{y}}^{-1}.
\end{align}
When $L_x$ and $L_y$ are both multiples of $p$, we can extend the product to cover the entire lattice,
\begin{align}
\prod_{j_x=1}^{L_x} \prod_{j_y=1}^{L_y} [b_{(1+j_x)\hat{x} + (1+j_y)\hat{y}}]^{j_xj_y} 
= M(X_0^{-1}).
\label{eq:px0-identity}
\end{align}
Since the ground states obey $b_i\ket{{\rm GS}}=\ket{{\rm GS}}$, Eq. (\ref{eq:px0-identity}) shows that $M(X_0)^{-1} |{\rm GS} \rangle = |{\rm GS} \rangle$ and hence $M(X_0 ) |{\rm GS} \rangle = |{\rm GS} \rangle$ fails to generate new ground states. However, it is impossible to express $M(X_0)$ in terms of the product of $b_i$'s as above when either $L_x$ or $L_y$ is not a multiple of $p$, and $M(X_0)$ becomes a true logical operator. The number of $M(X_0)$ logical operator is $1 -  \delta_{L_x {\rm mod}~p , 0} \delta_{L_y {\rm mod}~p , 0}$.

The next type of logical operators to consider are $H_{j_y} (X_1)$ and $V_{j_x} (X_2)$. In counting the number of independent $H_{j_y} (X_1)$'s, we invoke the identity
\begin{align}
\prod_{j_y=1}^p [H_{j_y}(b)]^{j_y} &=H_{0}(X_1)H_{p}(X_1^{-1}) 
\label{eq:x-ident-h2}
\end{align}
showing that $H_{j_y} (X_1 ) |{\rm GS} \rangle = H_{j_y + p} (X_1 ) |{\rm GS} \rangle$. For $L_y \md p \neq 0$, continued application of such equivalence relation implies $H_{j_y} (X_1 ) |{\rm GS} \rangle = H_{j'_y} (X_1 ) |{\rm GS} \rangle$ for all $j'_y$, leading to only one independent logical operator $H_{j_y} (X_1)$. 

For $L_y \md p = 0$, such argument breaks down and one can have as many as $p$ independent logical operators. However, one must deal with another identity
\begin{align}
H_{j_y} (b)&= H_{j_y -1}(X_1) H_{j_y} (X_1^{-2}) H_{j_y +1 } (X_1)  \label{eq:x-ident-h1}
\end{align}
which holds regardless of the system size. Since $H_{j_y} (b) |{\rm GS} \rangle = |{\rm GS} \rangle$,  the l.h.s. of the identity acting on the ground state is the same ground state, the action of $H_{j_y + 2} (X_1)$ on the ground state is equivalent to the consecutive actions of $H_{j_y} (X_1)$ and $H_{j_y +1} (X_1)$ on the same state. There are then at most two logical operators of this type. Since there is no identity relation  
connecting $H_{j_y} (X_1)$ to the neighboring operator $H_{j_y+1} (X_1)$, it can be conclude that there are actually two independent logical operators of the type $H_{j_y} (X_1)$ when $L_x\md p=0$. The number of independent horizontal logical operators can be summarized as $1+\delta_{0,L_y \md p}$. Following a similar procedure, the number of vertical logical operators $V_{j_x} (X_2)$ is $1+\delta_{0,L_x \text{mod} ~p}$. 

Finally, we have two types of logical operators ${\mathbb D}_i^{xy}$ and ${\mathbb D}_i^{x\bar{y}}$ along the diagonal directions $(1,1)$ and $(1,-1)$, defined as
\begin{align}
    {\mathbb D}_i^{xy} & \equiv D_i^{xy}(X_0) D_i^{xy}(X_1) D_i^{xy}(X_2), \nn
    {\mathbb D}_i^{x\bar{y}} & \equiv D_{i-\hat{y}}^{x\bar{y}}(X_0^{-1}) D_i^{x\bar{y}}(X_1) D_i^{x\bar{y}}(X_2), \label{eq:diag-X}
\end{align}
irrespective of $L_x$ and $L_y$. The diagonal operators appearing on the r.h.s., namely $D_i^{xy}(X_0)$, $D_i^{xy}(X_1)$, and $D_i^{xy}(X_0)$, do not commute with the Hamiltonian individually but their product as given above does. There are $L_g=\text{gcd}(L_x,L_y)$ distinct operators one can write down for each of the diagonal operators on the r.h.s. (see discussion in Sec. \ref{sec:5}), which result in $L_g$ distinct diagonal operators $\mathbb{D}_i^{xy}$ and $\mathbb{D}_i^{x\bar{y}}$ on the l.h.s. of Eq. (\ref{eq:diag-X}). Most importantly, ${\mathbb D}_i^{xy}$ and ${\mathbb D}_i^{x\bar{y}}$ do commute with the Hamiltonian, making them eligible as logical operators. 

Now one must count how many of the ${\mathbb D}_i^{xy}$ and ${\mathbb D}_i^{x\bar{y}}$ operators are truly independent. For that, we start with the identity 
\begin{align}
    \prod_{j_y=1}^p \left[D^{xy}_{i+j_y\hat{y}}(b)\right]^{j_y}   & = \mathbb{D}_i^{xy} \left[\mathbb{D}_{i+p\hat{y}}^{xy}\right]^{-1}, \nn
    \prod_{j_y=1}^p \left[D^{x\bar{y}}_{i+j_y\hat{y}}(b)\right]^{j_y} & = \mathbb{D}_i^{x\bar{y}} \left[\mathbb{D}_{i+p\hat{y}}^{x\bar{y}}\right]^{-1} ,
\end{align}
similar to Eq. (\ref{eq:x-ident-h2}) for the horizontal logical operators. The first identity shows that ${\mathbb D}_i^{xy}|{\rm GS}\rangle = {\mathbb D}_{i+p\hat{y}}^{xy}|{\rm GS}\rangle$. For $L_g \md p \neq 0$, then, every diagonal operator ${\mathbb D}_i^{xy}$ becomes equivalent, leaving only one independent logical operator $\mathbb{D}_i^{xy}$. Similarly, there is only one independent $\mathbb{D}_i^{x\bar{y}}$ when $L_g \md p \neq 0$. 

The argument breaks down for $L_g \md p = 0$, and one must consider another kind of identities
\begin{align}
    D_i^{xy}(b) & = \mathbb{D}_{i-\hat{y}}^{xy} \left[ \mathbb{D}_{i}^{xy}\right]^{-2} \mathbb{D}_{i+\hat{y}}^{xy}, \nn
    D_i^{x\bar{y}}(b) & = \mathbb{D}_{i-\hat{y}}^{x\bar{y}} \left[\mathbb{D}_{i}^{x\bar{y}}\right]^{-2} \mathbb{D}_{i+\hat{y}}^{x\bar{y}} . \label{eq:6.12}
\end{align}
Since $D_i^{xy} (b) |{\rm GS} \rangle = D_i^{x\bar{y}} (b) |{\rm GS} \rangle = |{\rm GS} \rangle$, one can express one of the diagonal operators on the r.h.s. of Eq. (\ref{eq:6.12}) in terms of the other two. There are thus only two independent logical operators among $\mathbb{D}_{i}^{xy}$'s and also among $\mathbb{D}_{i}^{x\bar{y}}$'s when $L_g \md p=0$.  

Since $p$ is a prime number, the condition $L_g ~{\rm mod}~p=0$ is equivalent to $L_x ~{\rm mod}~p = L_y~ {\rm mod}~p = 0 $. It then seems the number of independent $\mathbb{D}_i^{xy}$ and $\mathbb{D}_i^{x\bar{y}}$ operators are $2 + 2\delta_{0, L_x~ {\rm mod}~p}\delta_{0, L_y ~{\rm mod}~p}$. However, there exist additional identities constraining the independence of logical operators: 
\begin{align}
M(X_0) \prod_{j_y=1}^{L_y} H_{j_y}(X_1) \prod_{j_x=1}^{L_x}V_{j_x}(X_2)  & = \prod_{j_y=1}^{L_g}\mathbb{D}_{j_y\hat{y}}^{xy},\nn
M(X_0^{-1}) \prod_{j_y=1}^{L_y} H_{j_y}(X_1) \prod_{j_x=1}^{L_x}V_{j_x}(X_2)  & = \prod_{j_y = 1}^{L_g}\mathbb{D}_{j_y\hat{y}}^{x\bar{y}} . \label{eq:6.13} 
\end{align}
These identities apply irrespective of $L_x$ and $L_y$. $H_{j_y} (X_1)$ and $V_{j_x} (X_2)$ are valid logical operators for all values of $L_x , L_y$. According to the argument given at the beginning of the subsection, $M(X_0)$ is a logical operator when $L_x$ or  $L_y$ is incommensurate with $p$, but acts as an identity operator when both $L_x , L_y$ are commensurate with $p$. In either case, the two relations given in Eq. (\ref{eq:6.13}) place constraints on $\mathbb{D}_i^{xy}$ and $\mathbb{D}_i^{x\bar{y}}$ and reduce their degrees of freedom each by one, making the total number of diagonal operators equal to $2 \delta_{0, L_x~ {\rm mod}~p}\delta_{0, L_y ~{\rm mod}~p}$.

Gathering all the statements, we once again arrive at the number of logical operators given by $N_{lo} = 2 + \left( 1+ \delta_{0,L_x ~{\rm mod}~p}\right)$ $\left(1 + \delta_{0,L_y~ {\rm mod}~p} \right)$. 

\section{Excitations}
\label{sec:7}
The qusiparticle excitations in the R2TC model are monopoles and dipoles. The monopole quasiparticles can be efficiently characterized with the help of the following operators
\begin{align}
\mathbb{A}_i^{x}(j)  & = \frac{1}{p} \sum_{k=0}^{p-1} (\omega^{-j} a_i^x)^k, \nn
\mathbb{A}_i^{y}(j)  & = \frac{1}{p} \sum_{k=0}^{p-1} (\omega^{-j} a_i^y)^k, \nn
\mathbb{B}_{i}(j) & = \frac{1}{p} \sum_{k=0}^{p-1} (\omega^{-j} b_i)^k,
\label{eq:excitation-num}
\end{align}
where $j \in \mathbb{Z}$. They are defined in such a way that the eigenstates of $a_i^x ,~ a_i^y,~ b_i$ with eigenvalues equal to $\omega^{j}$ are automatically the eigenstates of $\mathbb{A}_i^x(j)$, $\mathbb{A}_i^y(j)$, and $\mathbb{B}_i (j)$, respectively, with the eigenvalue $+1$. The integer $j$ (mod $p$) serves as the charge of the monopole excitation at the site $i$. We use $p_x$, $p_y$, and $q$ in subsequent discussions to denote the monopole charges associated with $\mathbb{A}_i^x (p_x) $, $\mathbb{A}_i^y (p_y)$, and $\mathbb{B}_i (q)$, respectively. We will also refer to them as $\mathbb{A}^x, \mathbb{A}^y$, and $\mathbb{B}$ excitations. 

\begin{figure}[h]
\includegraphics[width=0.45\textwidth]{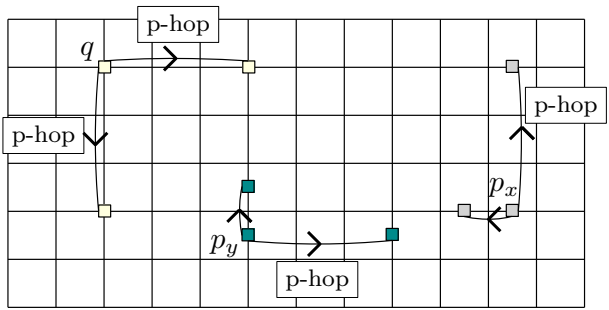}
\caption{Monopole quasiparticles in R2TC $(q, p_x , p_y)$. The $p$-hop process refers to hopping by $p$ lattice spacing. The $q$ quasiparticle can only execute $p$-hops in both $x$ and $y$ directions. The $p_x$ ($p_y$) monopole hops by one lattice spacing in the $x$ ($y$) direction and by $p$ spacing in the $y$ ($x$) direction. The $p$-hop processes are forbidden in the parent rank-2 U(1) LGT but are allowed now due to the relaxed constraint provided by Higgsing.}
\label{fig:monopole-motion}
\end{figure}

An intuitive way to think about the $q,~ p_x ,~ p_y$ quasiparticles in the R2TC is to view them as Higgsed version of the fracton ($\rho_i$) and lineon ($\rho_i^x , \rho_i^y$) excitations in the rank-2 U(1) LGT. The strict constraint on the fracton or the lineon motion becomes relaxed into $p$-hops in the previously forbidden direction thanks to the Higgsing process. The allowed motions of quasiparticles in R2TC are summarized graphically in Fig. \ref{fig:monopole-motion}.

The other type of quasiparticle excitation is the dipole consisting of two monopoles of opposite charges. For instance, two $\mathbb{B}_i$ monopoles with charges $\pm q$ separated by $\vec{d}=d_x\hat{x}+d_y\hat{y}$ and carrying the dipole moment $q \vec{d}$ is denoted $\Delta_{q,\vec{d}}$. Similarly, $\Delta_{p_x, \vec{d}}$ $(\Delta_{p_y, \vec{d}})$ refers to the dipole consisting of two $\mathbb{A}_i^x$ ($\mathbb{A}_i^y$) monopoles with charges $\pm p_x$ ($\pm p_y$) and the dipole moment $p_x\vec{d}$ ($p_y\vec{d}$). All these dipoles are free to move in any direction by one lattice spacing at a time. Recall that the dipoles in the parent LGT were also free. 

In the parent rank-2 U(1) LGT, dipoles are always created as a dipole and an anti-dipole pair. While such quadrupole creation is possible in the R2TC, there is the additional possibility of creating a single dipole without the accompanying pair. They are referred to as pristine (single dipole creation) and emergent (dipole pair creation) dipoles. Their respective creation processes will be discussed in the following subsections. 

\begin{figure*}[tb]
\includegraphics[width=0.83\textwidth]{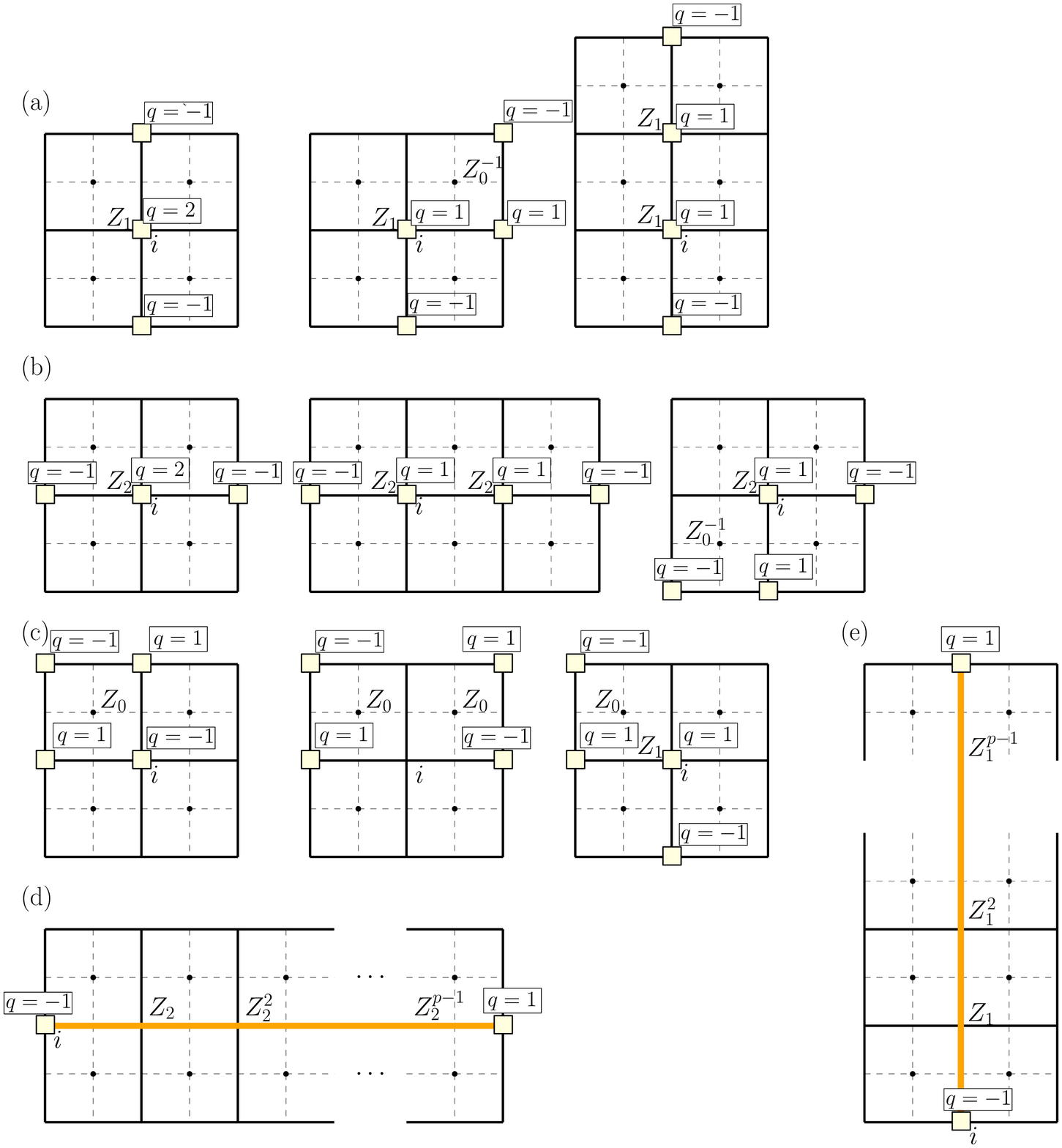}
\caption{Graphical illustration of the $\mathbb{B}$ excitations. (a) Applying a $Z_1$ operator to the ground state creates two $q=-1$ and one $q=2$ excitations along the vertical direction. Ways to move a dipole horizontally or vertically through the application of $Z_0^{-1}$ or $Z_1$ operator are illustrated in the subsequent panels.  (b) Applying a $Z_2$ operator creates two $q=-1$ and one $q=2$ excitations along the horizontal direction. Ways to move a dipole horizontally and vertically are illustrated. (c) Applying a $Z_0$ operator creates two $q=-1$ and two $q=1$ excitations around a plaquette. Ways to move a dipole after their creation are illustrated. (d) Pristine dipole creation operators $\mathbb{Z}_{x}$ and (e) $\mathbb{Z}_{y,i}$ are shown as orange lines with charges at either ends.}
\label{fig:B}
\end{figure*}

\subsection{$\mathbb{B}$ Excitations}
\label{sec:q-excitation}

The $\mathbb{B}$ excitations refer to eigenstates of $\mathbb{B}_i (j)$ for $j \neq 0$, characterized by nonzero charges $q$. For instance, the action of $Z_{1,i}$ or $Z_{2,i}$ operator on the ground state results in the creation of two $q = -1$ monopoles and one $q=2$ monopole along either vertical or horizontal direction, altogether forming a quadrupole excitation as shown in the leftmost panels of Fig. \ref{fig:B}(a) and (b). On the other hand, the action of $Z_{0,i}$ on a plaquette creates a quadrupole with two $q=-1$ charges and two $q=1$ charges placed at the four vertices of a plaquette as shown in the leftmost panel of Fig. \ref{fig:B}(c). One can view a quadrupole as a pair of emergent dipoles. 

Each emergent dipole can subsequently be moved through the lattice. For instance, the middle (rightmost) panel of Fig. \ref{fig:B}(a) shows the horizontal (vertical) displacement of an emergent dipole by the action of quadrupole creation operator $Z_{0,i}^{-1}$ ($Z_{1,i+\hat{y}}$) on an existing quadrupole. In essence, the quadrupole creation operator acts also as the hopping operator of a dipole. Similarly, the emergent dipole created in the leftmost panel of Fig. \ref{fig:B}(b) can move horizontally (vertically) through the action of the quadrupole operator $Z_{2,i+\hat{x}}$ [middle panel] ($Z_{0,i}$ [rightmost panel]). Finally, the emergent dipole movement in Fig. \ref{fig:B}(c) is executed horizontally by $Z_0$ and vertically by $Z_1$, as shown in the middle and rightmost panels. In conclusion, emergent dipoles 
can freely execute the {\it center-of-mass} movement after their creation by judicious application of the quadrupole opreators $Z_0$, $Z_1$, and $Z_2$. The same cannot be said of the {\it rotation} of emergent dipoles.

The pristine dipoles can be created singly without its partner. Figure \ref{fig:B}(d) illustrates the action of the $p$-local operator (meaning it acts on $p$ consecutive sites at once)
\begin{equation} 
\mathbb{Z}_{x,i} =\prod_{j_x=1}^{p} \bigl[ Z_{2,(i_x+j_x)\hat{x} + i_y \hat{y} } \bigr]^{j_x}\label{eq:Zx-operation2} 
\end{equation}
giving rise to one $q=1$ and one $q=-1$ excitations with the lattice spacing $p$ between them, but no other accompanying dipole. This is the creation operator of a pristine dipole. Similarly,
\begin{equation} \mathbb{Z}_{y,i} =\prod_{j_y=1}^{p} [ Z_{1,i_x\hat{x}+(i_y+j_y)\hat{y}} ]^{j_y} \label{eq:Zy-operation}  \end{equation}
creates a vertically oriented pristine dipole with the lattice spacing of $p$ between the $q=\pm 1$ charges as shown in Fig. \ref{fig:B}(e). The net dipole moment is $p\hat{x}$ or $p\hat{y}$, equivalent to 0 under the mod $p$ consideration. This is, in essence, why one can create such a dipole singly, since the dipole moment associated with it is zero mod $p$. The pristine dipoles are illustrated as orange lines extending over the $p$ lattice sites in Fig. \ref{fig:B}(d) and (e). 

Suppose we act with $\mathbb{Z}_{x,i}$ or $\mathbb{Z}_{y,i}$ on an existing dipole. The net effect is the hopping of a charge $q=1$ at one end of a dipole by the lattice spacing $p$ in either horizontal or vertical direction. As a result, a single $q$ monopole can move either horizontally and vertically through the action of $\mathbb{Z}_x$ or $\mathbb{Z}_y$, but only by $p$ lattice spacing at a time. The pristine dipole creation operators act as monopole hopping operators by $p$ lattice spacing.

It is well known that logical operators in the toric code have physical interpretation as the creation of an anyon pair, followed by their re-annihilation after one of the anyons is moved round the entire circumference of the torus. A similar interpretation applies to the logical operators we have identified in the previous section. For example, after the quadrupole creation shown in the leftmost panel of Fig. \ref{fig:B}(a), one of the dipoles can be moved vertically around the circumference of the torus and be annihilated with the remaining dipole. The operator that performs this operation is precisely $V_{j_x}(Z_1)$ - one of the logical operators identified in the previous section. Likewise, the quadrupole configuration created in the leftmost panel of Fig. \ref{fig:B}(b) will be re-annihilated by the action of another logical operator, $H_{j_y}(Z_2)$. Finally, the quadrupole configuration in the leftmost panel of Fig. \ref{fig:B}(c) is annihilated by $H_{j_y}(Z_0)$ or $V_{j_x}(Z_0)$. These four logical operators are the analogues of the logical operators in the toric code, with the difference that now the operators move the dipoles rather than the monopoles (anyons). The last two operators, $H_{j_y}(Z_0)$ and $V_{j_x}(Z_0)$, are actually dependent on each other since they both are related with $M(Z_0)$ as pointed out in Eq. (\ref{eq:Z0-identity1}). Hence, in this case, we have not four, but three linearly independent logical operators doing the quadrupole creation and re-annihilation. 

The creation and re-annihilation of a pristine dipole leads to another set of logical operators. As mentioned, a pristine dipole can be created by the action of the $p$-local operators in Eqs. (\ref{eq:Zx-operation2}) and (\ref{eq:Zy-operation}). Logical operators responsible for the creation and re-annihilation of a single pristine dipole must be able to move a monopole round the circumference of the torus and annihilate it with the remaining monopole pair. The operators for these processes are precisely $\mathbb{H}_{j_y}(Z_2)$ and $\mathbb{V}_{j_x}(Z_1)$ defined in Eq. (\ref{eq:big-H-big-Z}). One can see that these operators move a monopole by $p$ lattice spacings at a time. In order to encounter the partner monopole, the linear dimension $L_x$ or $L_y$ needs to be a multiple of $p$, otherwise the monopole hop would have to take place $p$ times round the torus in order to finally meet its partner. That, however, requires taking the $p$-th power of $\mathbb{H}_{j_y}(Z_2)$ or $\mathbb{V}_{j_x}(Z_1)$ which is a trivial identity and not a logical operator. To conclude, logical operators associated with the creation and re-annihilation of pristine dipoles only exist if either $L_x$ or $L_y$ is commensurate with $p$. This is the physical picture as to why the number of logical operators and hence the GSD depends on the mod $p$ parity of $L_x$ and $L_y$. 

For both $L_x$ and $L_y$ commensurate with $p$, the two quadrupole operators $H_{j_y}(Z_0)$ and $V_{j_x}(Z_0)$ become independent operators as the constraint (\ref{eq:Z0-identity1}) is no longer meaningful. In this case, we end up having six logical operators: four quadrupole operators,  $H_{j_y}(Z_0),~ V_{j_x}(Z_0),~ H_{j_y}(Z_2),~ V_{j_x}(Z_1)$, and two dipole operators $\mathbb{H}_{j_y}(Z_2) , \mathbb{V}_{j_x}(Z_1)$. Each logical operator contributes to the degeneracy $p$ of the ground state, hence, we have $p^6$ GSD. This result is consistent with the conclusion in Ref. \onlinecite{barkeshli18}, where the GSD is given by $2^6$ with $p=2$ for even values of $L_x$ and $L_y$. The interpretation given by these authors is that the model is three copies of $\mathbb{Z}_p$ toric code when $p=2$ and each toric code has its own horizontal and vertical logical operators. Although their interpretation is insightful, it cannot cover the case where $L_x$ or $L_y$ is not commensurate with $p$. According to our interpretation, four of the logical operators are rightfully interpreted as quadrupolar, and only two are dipolar. This interpretation fully covers the case where $L_x$ or $L_y$ is incommensurate with $p$. 

\subsection{$( \mathbb{A}^x , \mathbb{A}^y )$ Excitations}
\begin{figure}[tb]
\includegraphics[width=0.46\textwidth]{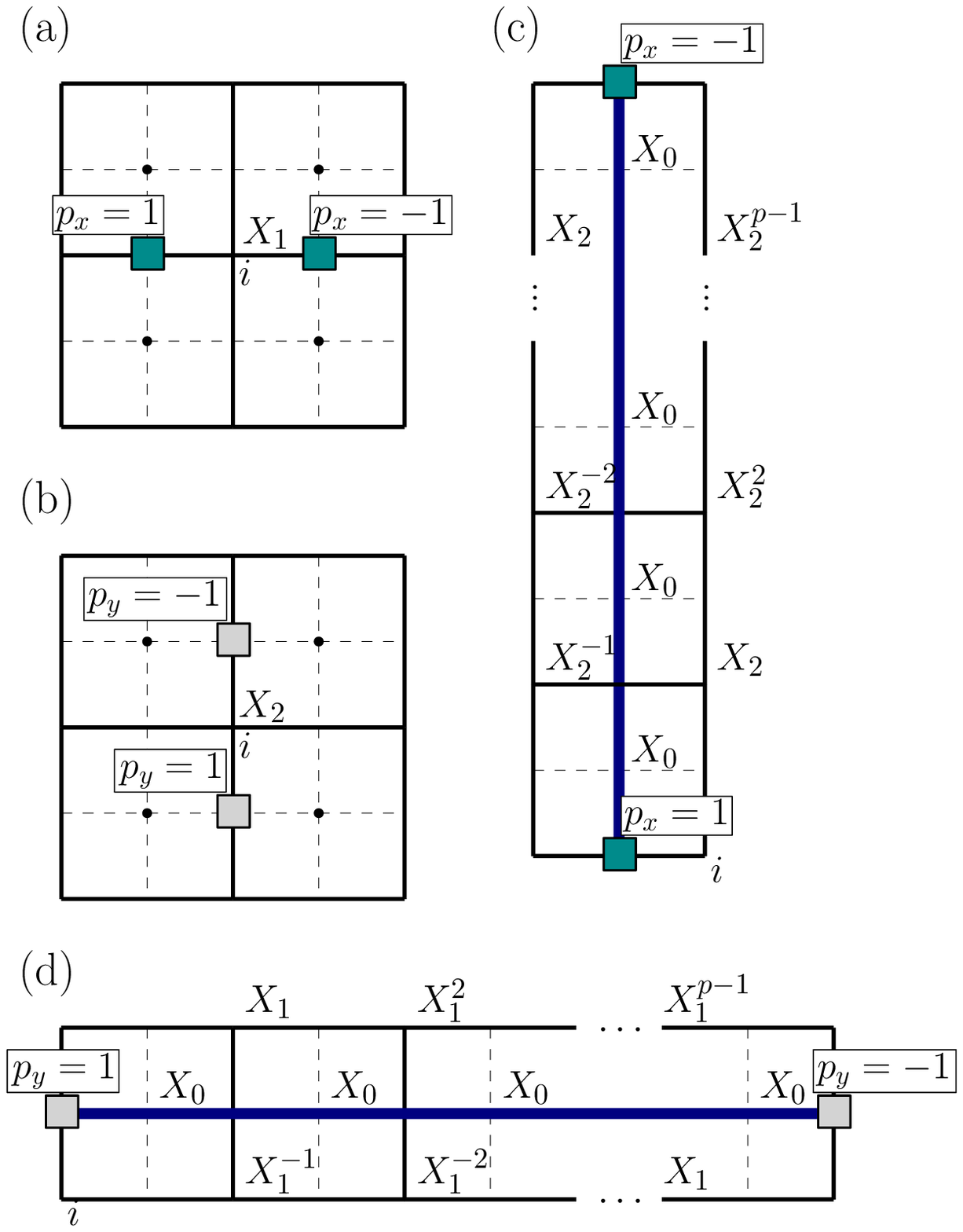}
\caption{Graphical illustration of the $\mathbb{A}^x$ (cyan square) and $\mathbb{A}^y$ (gray square) excitations. (a), (b) Pristine dipole creation operators $X_{1,i}$ and $X_{2,i}$ are illustrated. 
(c), (d) Pristine dipoles created by $\mathbb{X}_{y,i}$ and  $\mathbb{X}_{x,i}$ are shown as navy lines.}
\label{fig:a}
\end{figure}

While the $\mathbb{B}_i$ excitations were best viewed as residing at the sites, the $(\mathbb{A}_i^x , \mathbb{A}_i^y )$ excitations are best viewed as those at the links. They are denoted as cyan and gray squares at the corresponding links in Fig. \ref{fig:a}. In the case of $\mathbb{B}$ excitations, pristine dipoles were always of length $p$. This is no longer the case for $(\mathbb{A}^x , \mathbb{A}^y)$ pristine dipoles, which can be either of length 1 or $p$. 

The pristine $p_x$ ($p_y$) dipole consisting of a pair of $\mathbb{A}_i^x$ ($\mathbb{A}_i^y$) excitations in the horizontal (vertical) direction can be formed with any length. By applying $X_{1,i}$ ($X_{2,i}$) on the ground state, we can create a pristine dipole consisting of $p_x =\pm1$ ($p_y =\pm1$) separated by one lattice site in horizontal (vertical) direction of $\mathbb{A}_i^x$ ($\mathbb{A}_i^y$), as shown in Fig. \ref{fig:a}(a) (Fig. \ref{fig:a}(b)). Since $X_1$ and $X_2$ are pristine dipole creation operators, when we act $X_1$($X_2$) on an existing $p_x$($p_y$) dipole, a $p_x=-1$($p_y=-1$) monopole will hop by $1$ lattice spacing horizontally(vertically). By repeating the operation, the length of the pristine dipole can be arbitrarily large. Moving the charge in a pristine dipole in the direction orthogonal to the dipole moment will create an additional dipole and increase energy.

\begin{figure}[tb]
\includegraphics[width=0.48\textwidth]{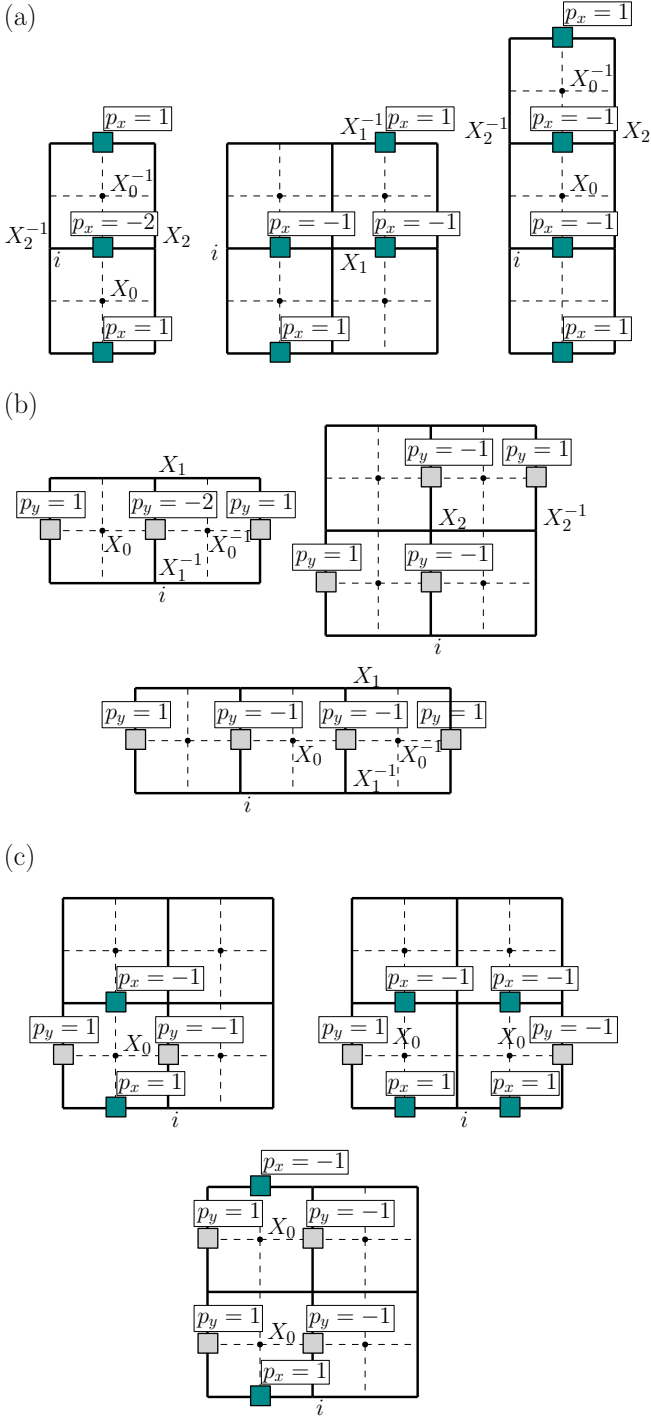}
\caption{Pair creations of (a) $\Delta_{p_x, \hat{y}}$ and (b) $\Delta_{p_y,\hat{x}}$, and their subsequent horizontal and vertical motions. (c) Illustration of various actions of $X_0$. See main text for details.}
\label{fig:a2}
\end{figure}

The pristine $p_x$ ($p_y$) dipole can be created along the vertical (horizontal) direction as well, but it is only of length $p$. The creation operators of such dipoles are 
\begin{widetext}
\begin{align}
\mathbb{X}_{x,i}=\prod_{j_x=1}^p &\Bigl[X_{0,{(i_x+j_x-1)\hat{x}+(i_y-1)\hat{y}}} X_{1,(i_x+j_x)\hat{x}+i_y\hat{y}}^{-j_x} X_{2,(i_x+j_x)\hat{x}+(i_y+1)\hat{y}}^{j_x}\Bigl] \nn
\mathbb{X}_{y,i}=\prod_{j_y=1}^p %
&\Bigl[X_{0,{(i_x-1)\hat{x}+(i_y+j_y-1)\hat{y}}} X_{2,(i_x-1)\hat{x}+(i_y+j_y)\hat{y}}^{-j_y} X_{2,i_x\hat{x}+(i_y+j_y)\hat{y}}^{j_y} \Bigl] . \label{eq:XxXy} 
\end{align}
\end{widetext}
Applying $\mathbb{X}_{y,i}$ ($\mathbb{X}_{x,i}$) create a $p_x$($p_y$) pristine dipole of length $p$ vertically [Fig. \ref{fig:a}(c)] (horizontally [Fig. \ref{fig:a}(d)].

The pristine $p_x$ ($p_y$) dipole creation operators act as hopping operators of $p_x$ ($p_y$) monopoles. The $X_1$ and $\mathbb{X}_y$ operators are responsible for the $\mathbb{A}^x$ monopole movement in the horizontal movement by one spacing and the vertical movement by $p$ spacing, respectively. Similarly, $X_2$ and $\mathbb{X}_x$ operators move the $p_y$ monopole by one lattice spacing vertically and $p$ lattice spacing horizontally. 

The quadrupole creation operators create emergent $p_x$ or $p_y$ dipoles in pairs. They are defined as 
\begin{align}
\mathcal{X}^{p_x,\hat{y}}_{y,i} =& X_{0,i_x \hat{x}+(i_y-1)\hat{y}} X_{0,i_x\hat{x} + i_y\hat{y}}^{-1} X_{2,i_x\hat{x} + i_y\hat{y}}^{-1} X_{2,(i_x+1)\hat{x} + i_y \hat{y}}\nn
\mathcal{X}^{p_y,\hat{x}}_{x,i} =& X_{0,(i_x-1) \hat{x}+i_y\hat{y}} X_{0,i_x\hat{x} + i_y\hat{y}}^{-1} X_{1,i_x\hat{x} + i_y\hat{y}}^{-1} X_{1,i_x\hat{x} + (i_y+1) \hat{y}}. \label{eq:pxpy-dipole-pair-creation}
\end{align}
As depicted in the leftmost panels of Fig. \ref{fig:a2}(a) and (b), they create a pair of dipoles $\Delta_{p_x, \hat{y}} = \pm \hat{y}$ separated vertically by one lattice spacing, and $\Delta_{p_y, \hat{x}} = \pm \hat{x}$ dipole pair separated horizontally by one lattice spacing, respectively. 

One can move the emergent dipole $\Delta_{p_x, \hat{y}}$ in the vertical direction by applying the dipole pair creation operator $\mathcal{X}^{p_x,\hat{y}}_{y,i}$ in succession, as shown in the rightmost panel of Fig. \ref{fig:a2}(a). 
As shown in the middle panel of Fig. \ref{fig:a2}(a), the horizontal motions of $\Delta_{p_x, \hat{y}}$ are implemented by applying $X_1$ or $X_1^{-1}$. Similarly, the horizontal movement of the $\Delta_{p_y,\hat{x}}$ dipole is done by applying the operator $\mathcal{X}^{p_y,\hat{x}}_{x,i}$ in succession, as shown in the bottom panel of Fig. \ref{fig:a2}(b). Their vertical motions are implemented by applying $X_2$ or $X_2^{-1}$. 

Whereas the operators in Eq. (\ref{eq:pxpy-dipole-pair-creation}) create quadrupoles consisting of two $p_x$ dipoles or two $p_y$ dipoles, the action by $X_{0,i}$ on the ground state creates a quadrupole consisting of one $p_x$ dipole and one $p_y$ dipole,
as shown in the top left panel of Fig. \ref{fig:a2}(c). Applying $X_1$($X_2$) on a $p_x$($p_y$) monopole with charge $1$($-1$) in this configuration creates a horizontal(vertical) monopole movement as explained previously. It should be kept in mind that the continued operation of $X_{0,i}$'s will increase the number of $p_x$ and $p_y$ monopoles and the energy as well. Applying $X_{0,i}$ on the quadrupole configuration shown in the top left panel of Fig. \ref{fig:a2}(c) creates additional $p_x$ dipoles as illustrated in the top right panel of Fig. \ref{fig:a2}(c). Similarly, applying $X_{0,i-\hat{x}+\hat{y}}$ creates a $p_y$ dipole in its wake as shown in the the bottom panel of Fig. \ref{fig:a2}(c). 

\section{Braiding Statistics}
\label{sec:8}

Various ways to move the monopole and dipole excitations after their creation were discussed in the previous section. Based on the knowledge one can move one quasiparticle around the other and calculate the statistical phase resulting from such braiding. The phases of braiding statistics among the monopoles and the dipoles are summarized in Table \ref{tab:statist}.

We start with the case of monopole-monopole braiding. To do that, one must prepare two monopoles to participate in the braiding. This is done by first creating two pristine dipoles, and then pushing one of the monopoles from each dipole out to infinity, leaving a pair of monopoles for braiding. There are three types of monopoles, labeled $(q, p_x , p_y)$, but it turns out only the $(q, p_x)$ and $(q,p_y)$ monopole pairs yield nontrivial braiding phases. This is because, at the crossing point of the path of two different monopoles, pristine dipole creation operators for both monopoles, responsible for the motions of each monopole, are acted. Then, the braiding phase originates from the exchange of the operators. Since all the operators responsible for the motion of $p_x$ and $p_y$ consist of $X$'s, their braiding phase is always trivial (+1) and the statistic is bosonic.

\begin{table}[h]
\begin{tabular}{ |c|c|c|c|c| }
\hline
& ~~~\multirow{2}{*}{$p_x$}~~~ & ~~~\multirow{2}{*}{$p_y$}~~~ & ~~~\multirow{2}{*}{$\Delta_{p_x,\vec{d}}$}~~~ & ~~~\multirow{2}{*}{$\Delta_{p_y,\vec{d}}$}~~~\\
& & & &\\
\hline 
~~~\multirow{2}{*}{$q$}~~~& \multirow{2}{*}{$\omega^{p_x q (y_x-y_q)}$} &\multirow{2}{*}{$\omega^{p_y q (x_q- x_y)}$} & \multirow{2}{*}{$\omega^{p_xqd_y}$} & \multirow{2}{*}{$\omega^{-p_yqd_x}$}\\
& & & & \\
\hline
\multirow{2}{*}{$\Delta_{q,\vec{d}}$}& \multirow{2}{*}{$\omega^{-p_xqd_y}$} &\multirow{2}{*}{$\omega^{p_yqd_x}$} & \multirow{2}{*}{$1$} & \multirow{2}{*}{$1$}\\
& & & &\\
\hline
\end{tabular}
\caption{Braiding statistics among monopoles $(q,~ p_x ,~ p_y)$ and dipoles ($\Delta_{q, \vec{d}},~ \Delta_{p_x, \vec{d}},~ \Delta_{p_y , \vec{d}})$.}
\label{tab:statist}
\end{table}

The unusual aspect of the monopole-monopole braiding phases, as shown in Table \ref{tab:statist} is their dependence on the relative separation as $\omega^{p_x q ( y_x-y_q )}$ or $\omega^{p_y q ( x_q-x_y )}$, where $(x_q, y_q)$,  $(x_x , y_x)$, $(x_y , y_y)$ refer to the initial positions of the $q$, $p_x$ and $p_y$ monopoles, respectively. The usual anyonic braiding, on the other hand, gives a path-independent statistical phase. It is important to point out that, despite the apparent dependence on the initial positions of the quasiparticles, the braiding phases retain some of the topological character in that changes in the initial coordinates of the monopoles can only occur in a manner that preserves the phase. Suppose a different initial position $y_x$ of the $p_x$ monopole had been chosen for calculating the braiding phase. Due to the constrained motion of the $p_x$ quasiparticle analyzed in the previous section, the new $y$-coordinate can differ from old only in multiples of $p$, meaning that the phase $\omega^{p_x q ( y_x-y_q )}$ remains invariant for the new initial coordinate. The invariance of $\omega^{p_y q ( x_q-x_y )}$ under the coordinate change can be argued in the same manner.

Table \ref{tab:statist} also works out the braiding phase between an emergent dipole and a monopole. They can be obtained readily by employing the results of the monopole-monopole braiding statistics since a dipole is nothing but a pair of monopoles separated by distance $d=|\vec{d}|$. Contrary to the monopole-monopole braiding, the monopole-dipole braiding phases have no dependence on the initial coordinates. In other words, the emergent dipoles have an {\it abelian anyonic} braiding character with respect to the monopoles. The braiding of a pristine dipole with a monopole, on the other hand, always results in the trivial phase +1, which endows it with the {\it bosonic} character.  

In the rest of this section, we examine the braiding statistics of $q$ monopole with respect to the $p_x$ monopole when the monopole charges are both +1. To obtain the statistical phase, one can either (\rom{1}) braid $p_x$ round $q$ or (\rom{2}) braid $q$ round a fixed $p_x$. Both will result in the same statistical phase.

\begin{figure*}[tb]
\includegraphics[width=0.8\textwidth]{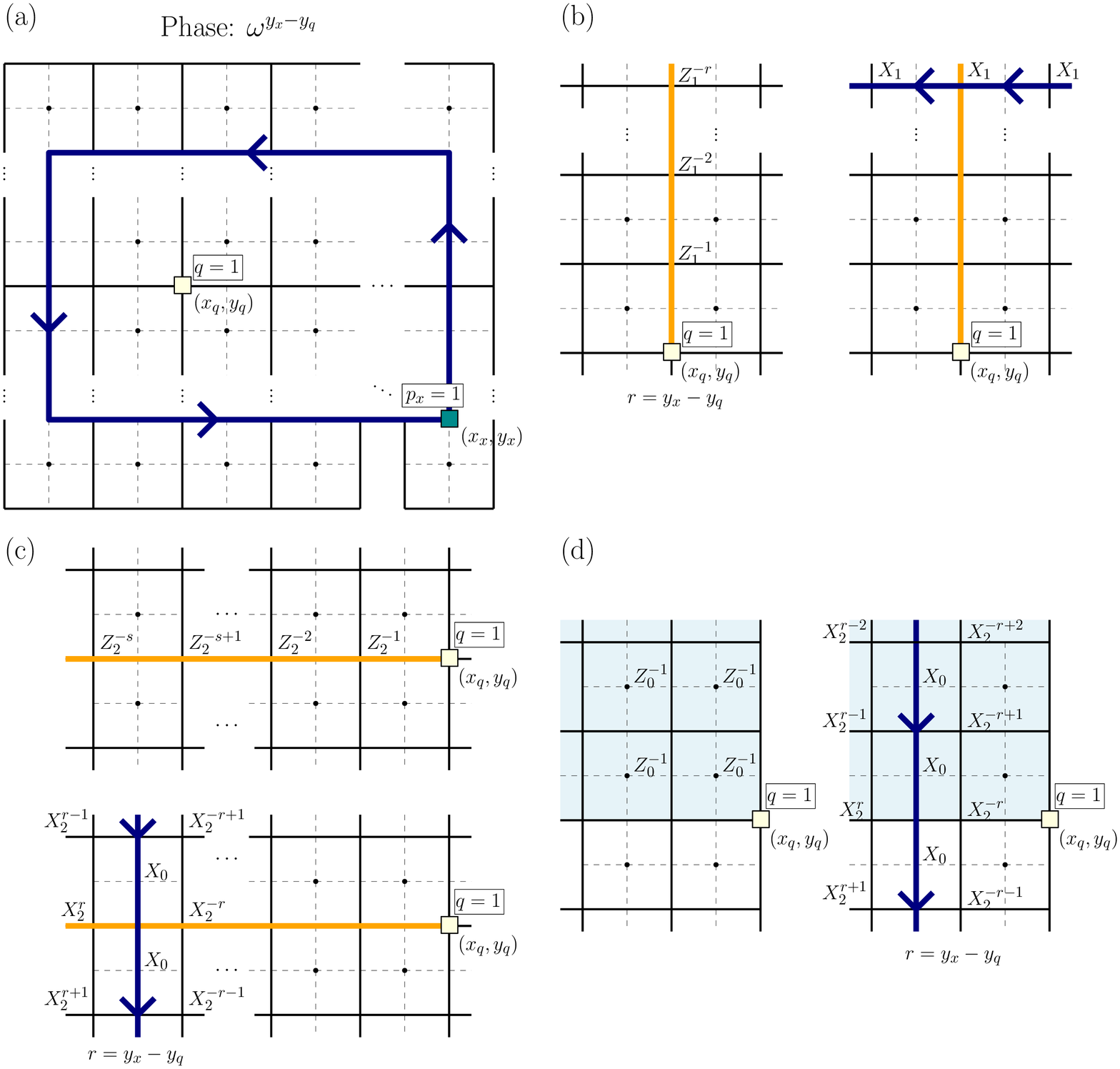}
\caption{Braiding of the $p_x=1$ monopole around a $q=1$ monopole. (a) A $p_x$ monopole winds around an isolated $q$ monopole counterclockwise. The $q$ monopole is fixed at $(x_q,y_q)$ while the $p_x$ monopole starts the braiding process from the position $(x_x,y_x)$. The braiding statistics is given by $\omega^{y_x - y_q}$. (b)-(d) Three ways of isolating the monopole $q$ are depicted in the first diagram of each figure. The orange lines represent the path in which the partner monopole is expelled to infinity. The braiding path of the $p_x$ monopole is shown by navy-colored lines. The blue membrane in (d) represents the $Z_0$ operation which pushes the other three monopoles of the quadrupole to infinity, leaving a single monopole $q$ in isolation.}
\label{fig:a_braid_b}
\end{figure*}

\subsection{Braiding $p_x$ round $q$}
\label{sec:px_braid_q}
In this subsection, we examine the statistical phase for braiding $p_x$ round $q$. As illustrated in Fig. \ref{fig:a_braid_b}(a), one can place an isolated $q$ monopole at $(x_q,y_q)$ and, starting from $(x_x,y_x)$, let a $p_x$ monopole travel counterclockwise around it. There are three ways to prepare an isolated $q$, all of them resulting in the same statistics. Although discussing all three methods in detail may seem redundant, we press forward with it for the sake of convincing the readers (and ourselves) that the rather peculiar statistical phase between the monopoles is indeed genuine to our model. 

The first way of isolating $q$ is illustrated in Fig. \ref{fig:a_braid_b}(b), where $q=1$ monopole is fixed at $(x_q, y_q)$ while its partner $q=-1$ is sent off to infinity along the orange-colored vertical line by the repeated operation of $\mathbb{Z}_y^{-1}$. Now, $p_x$ winds around $q$ counterclockwise along the navy-colored path in the right panel of Fig. \ref{fig:a_braid_b}(b). The hopping of $p_x$ in the $-x$ direction is implemented by applying $X_1$'s along the navy line. Starting at $(x_x , y_x)$, the $y$ coordinate of $p_x$ can change only in multiples of $p$. From the definition of the $\mathbb{Z}_y$ operator in Eq. (\ref{eq:Zy-operation}), we can see that the $Z_1$ operator on the orange line is raised to the power equal to the relative $y$ coordinate with respect to $y_q$. At the point where the orange and the navy line cross, $X_1$ encounters $Z_1^{-1}$ raised to the power $y_x - y_q$ at the point of intersection. Employing the commutation algebra of $X$ and $Z$ operators one gets
\begin{align}
X_1 Z_1^{y_q - y_x} & = \omega^{y_x - y_q} Z_1^{y_x - y_q} X_1.
\end{align}
The phase factor arising from the braiding of $p_x=1$ monopole around the $q = 1$ monopole is $\omega^{y_x-y_q}$. For other monopole charges, one merely adds multiplicative factors $p_x q$ and finds the expression listed in Table \ref{tab:statist}. 

The second way of isolating a $q=1$ monopole is illustrated in the top panel of Fig. \ref{fig:a_braid_b}(c), where a $q=1$ monopole is separated from its $q=-1$ partner through hopping of the latter along the horizontal orange line by repeated application of the $\mathbb{Z}_x^{-1}$ operator defined in Eq. (\ref{eq:Zx-operation2}). Afterwards, the $p_x = 1$ monopole moves along the navy line as in the bottom panel of Fig. \ref{fig:a_braid_b}(c) and traverses the orange line by the repeated implementation of $\mathbb{X}_y^{-1}$ defined in Eq. (\ref{eq:XxXy}). As explained earlier, the path of the $p_x$ monopole in the $y$ direction is best represented as a line on a dual lattice, while its path in the $x$ direction is best represented as a line on the links in the original square lattice [Figs. \ref{fig:a}(a) and (c)]. There is no restriction on the movement of the $p_x$ monopole in the $x$ direction, meaning that the $x$ coordinate of the sites at the intersection of the orange and the navy paths can be arbitrary. 

The $\mathbb{Z}_x^{-1}$ operator we exploited in isolating the $q$ monopole consists of the product of $Z_2$ operators raised to various powers along the orange line. It implies that the navy line representing the vertical path of the $p_x$ monopole can be positioned in the middle of two adjacent sites at which the $Z_2^{-1}$ operators in the orange line are raised to any successive integers, say $s$ and $s-1$. Meanwhile, the hopping of $p_x$ along the $-y$ direction is implemented by $\mathbb{X}_y$. Upon the examination of the definition of $\mathbb{X}_y$ we conclude that the power of $X_2$ on the left side of the navy line decreases by $1$ along the $y$ direction while it increases by $1$ on the right side of it. Therefore, at the sites located just to the left and the right of the crossing, the powers of $X_2$ become $r$ and $-r$, where $r=y_x-y_q$ is the $y$ coordinate difference of $p_x$ and $q$ at the start of the braiding. As a result, $X_2^{r}$ encounters $Z_2^{-s}$ and $X_2^{-r}$ encounters $Z_2^{-s+1}$ in the process of braiding. Such encounter produces the overall braiding statistics given by
\begin{align}
X_2^{r} Z_2^{-s} & = \omega^{rs} Z_2^{-s} X_2^{r}, \nn
X_2^{-r} Z_2^{-s+1} & = \omega^{-rs + r}  Z_2^{-s+1} X_2^{-r}.
\end{align}
The combined phase $\omega^{rs} \omega^{-rs + r}  = \omega^{r} = \omega^{y_x-y_q}$ is precisely the phase we obtained from the first procedure. 

Finally, a third way of isolating $q=1$ monopole is by the action of the quadrupole operator $Z_0^{-1}$ over a quadrant of the two-dimensional lattice with the monopole $q$ at its apex $(x_q , y_q)$, as illustrated in the left panel of Fig. \ref{fig:a_braid_b}(d). Operating with $Z_0^{-1}$ on all the plaquettes inside the quadrant pushes three of the constituent monopoles in a quadrupole to infinity, leaving behind one isolated monopole $q=1$. Afterwards, the $p_x$ monopole can braid the isolated $q$ by one horizontal movement and one vertical movement since only the movements of $p_x$ that take place inside the quadrant contribute to the braiding statistics while all other movements outside the quadrant do not. The $x$ movement of the $p_x$ monopole is carried out by $X_1$, which commutes with all the $Z_0$'s in the quadrant. Therefore the only non-trivial contribution to the braiding statistics comes from the vertical movement of the $p_x$ monopole shown as navy-colored line in the right panel of Fig. \ref{fig:a_braid_b}(d), implemented by repeated application of $\mathbb{X}_y$. Equation (\ref{eq:XxXy})] shows that $\mathbb{X}_y$ involves $X_0$ in its definition and, by carefully counting how many powers of $X_0$ appear inside the quadrant, we arrive once again at the statistical phase found previously.

\begin{figure*}[tb]
\includegraphics[width=0.8\textwidth]{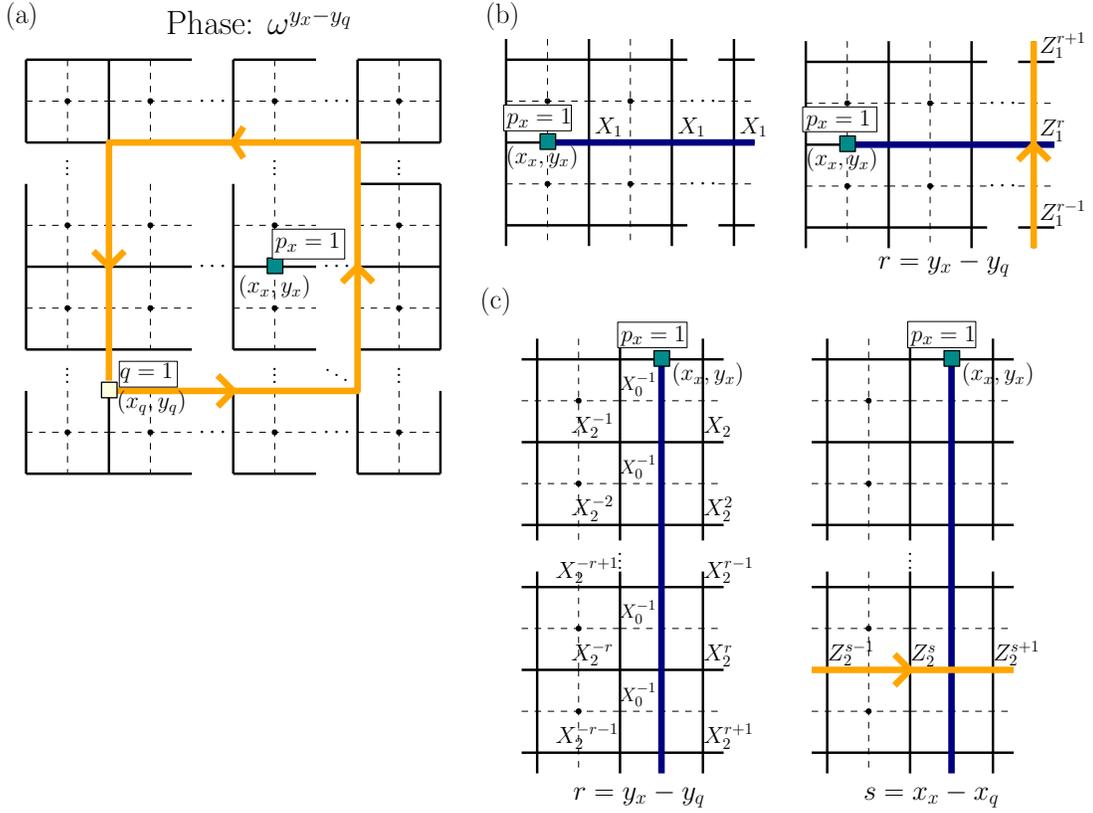}
\caption{Braiding of $q$ around $p_x$. (a) A $q=1$ monopole winds around an isolated monopole $p_x=1$, held fixed at $(x_x,y_x)$. The initial location of the $q$ monopole is $(x_q,y_q)$. The braiding phase equals $\omega^{y_x-y_q}$. Two different ways of isolating the $p_x =1$ monopoles are depicted in the first panels of (b) and (c). In both figures, the partner monopole is sent off to infinity along the navy-colored line. The $q$ monopole then winds around $p_x$ along the orange line. 
}
\label{fig:b_braid_a}
\end{figure*}

\subsection{Braiding $q$ round $p_x$}
In this subsection, we examine the statistical phase for braiding $q$ round $p_x$. Figure \ref{fig:b_braid_a}(a) shows the braiding of $q = 1$ monopole around the isolated $p_x = 1$ monopole in the counterclockwise direction. There are two ways to isolate the $p_x$ monopole. The first is by sending its dipolar partner $p_x = -1$ off to infinity through repeated operation of $X_1$ in the $x$ direction, along the path highlighted as navy line in the first row of Fig. \ref{fig:b_braid_a}(b). Afterwards, the $q = 1$ monopole winds around $p_x$ along the path depicted as orange-colored vertical line in the second row of Fig. \ref{fig:b_braid_a}(b). The propagation of $q$ by $p$ lattice spacing along the $y$ direction is implemented by repeated $\mathbb{Z}_y$ operation along the orange line. We have the operator $Z_1^{r}$ at the crossing point of the orange and the navy lines, where $r = y_x - y_q$ is the $y$ coordinate difference of the $p_x$ and $q$ monopoles at the start of braiding. This implies that $Z_1^{r}$ encounters $X_1$ at the crossing point, which yields the exchange operation
\begin{align*}
Z_1^{r} X_1 = \omega^{y_x - y_q}  X_1 Z_1^{r}.
\end{align*}

The second way of isolating $p_x=1$ is by applying  $\mathbb{X}_y^{-1}$ operator repeatedly along the navy-colored vertical line depicted in the left panel of Fig. \ref{fig:b_braid_a}(c). Afterwards, the braiding path of $q$ traverses the navy line as depicted with the orange line in the right panel of Fig. \ref{fig:b_braid_a}(c). The navy line is positioned at the middle of two adjacent sites of which the distances from $x_q$ are $s$ and $s+1$, where $s=x_q-x_x$. Note that the $x$-directed propagation of $q$ is implemented by the repeated operation of $\mathbb{Z}_x$ along the orange line, on which the power of $Z_2$ equals the relative $x$ coordinate with respect to $x_q$. Therefore, $Z_2^{s}$ and $Z_2^{s+1}$ are applied to the sites just to the left and right of the crossing point, respectively. In addition, during the propagation of $p_x$ along the $-y$ direction, the power of $X_2$ on the left side of the navy line decreases by $1$, while the power of $X_2$ on the right the navy line increases by $1$. As a result, at the sites located just to left and right of the crossing point, $Z_2^s$ encounters $X_2^{-r}$ and $Z_2^{s+1}$ encounters $X_2^{r}$, respectively, where $r = y_x - y_q$. Hence, we have two exchange operations for braiding statistics given by
\begin{align*}
    Z_2^s X_2^{-r} & = \omega^{-sr} X_2^{r} Z_2^s \nn 
    Z_2^{s+1} X_2^{r} & = \omega^{sr + r} X_2^{r}Z_2^{s+1}.
\end{align*}
The resulting combined phase is $\omega^{-sr} \omega^{sr -r} = \omega^{y_x -y_q}$. Following an entirely similar procedure, one finds the braiding phase $\omega^{x_q - x_y}$ between $q=1$ monopole and $p_y=1$ monopole when their initial positions are $(x_q,y_q)$ and $(x_y,y_y)$.


\section{Aharonov-Bohm phase}
\label{sec:9}
It must be admitted that the phase factors coming from the $(q, p_x)$ and $(q, p_y)$ braiding are highly unusual. In this section, we present some ways to understand these factors in the framework of the conventional Aharonov-Bohm (AB) phase. It is well known that the statistical phase can be understood as an AB phase due to the magnetic flux tightly attached to the quasiparticles. We can view the monopole-monopole braiding phases as also arising from the magnetic flux localized at the position of the monopole, which is experienced by the test monopole. We regard the monopole $q$ at ${\bm r}_q = (x_q , y_q)$ as the one with the magnetic flux attached and the $p_x$ or $p_y$ monopole as the test charge. The situation corresponds to $p_x$ braiding round $q$.

The magnetic flux attached to the monopole $q$ is given by Eq. (\ref{eq:b-field}) with $(m,n)=(1,1)$ which becomes, in the continuum limit, 
\begin{align}
B &= \frac{1}{2}(1+\delta_{b,d})\epsilon^{ab}\epsilon^{cd} \partial_a \partial_c A^{bd}\nn
&= \partial_x^2 A^{yy} + \partial_y^2 A^{xx} - \partial_x \partial_y A^{xy}.\label{eq:Bex}
\end{align}
The localized magnetic field $B({\bm r}) =( 2\pi q /p )  \delta^2({\bm r} - {\bm r}_q )$ in the rank-2 U(1) gauge theory comes from the tensor potential such as 
\begin{align}
A^{xx}({\bm r}) &= \frac{2\pi q}{p} (y-y_q) \delta(x - x_q) \theta(y-y_q) \nn
A^{xy}({\bm r}) &= A^{yy}({\bm r}) = 0 . \label{eq:Aab-for-localized-B}
\end{align}
One can indeed show that a localized magnetic flux is generated by this choice of the tensor potential using Eq. (\ref{eq:Bex}). Other choices of tensor potential for the same magnetic field are discussed in the Appendix.

It is argued that the $(q, p_x)$ and $(q, p_y)$ braiding phases $\phi_x$ and $\phi_y$ obtained in the previous section are captured as area integrals
\begin{align}
\phi_x & = p_x \int_{x_-}^{x_+} dx \int_{y_-}^{y_+} dy \left[ \int_{y_0}^y B (x,y') dy'  \right] \nn
\phi_y & = - p_y \int_{x_-}^{x_+} dx \int_{y_-}^{y_+} dy \left[ \int_{x_0}^x B (x' ,y) dx'  \right] . \label{eq:phixphiy} \end{align}
The integration area is assumed to be a square $[x_- , x_+ ] \times [y_- , y_+ ]$, which reflects the braiding path used in the previous section. The key difference from the conventional AB phase integral is that the $x$ or $y$ integral of the $B$-field is being used as an integrand, instead of the $B$ field itself. Performing the integrals assuming the localized magnetic field $B({\bm r}) = ( 2\pi q /p )  \delta^2({\bm r} - {\bm r}_q )$ gives 
\begin{align}
\phi_x & = \frac{2\pi q p_x}{p}  [ (y_+ \!-\! y_q ) - (y_+ \!-\!  y_- ) \theta (y_0 \! -\!  y_q ) ]  , \nn
\phi_y & = - \frac{2\pi q p_y}{p} [ (x_+ \!-\!  x_q ) + (x_+ \!-\!  x_- ) \theta (x_0 \! -\!  x_q ) ] . \label{eq:phixphiy2} 
\end{align}
Depending on the choice of the integration constant $y_0$, one gets $\theta(y_0 - y_q) =1$ and $\phi_x = (2\pi q p_x / p ) (y_- - y_q )$, or $\theta(y_0 - y_q ) = 0$ and $(2\pi q p_x /p ) (y_+ - y_q )$. Recall further that the $p_x$ monopole can only hop by $p$ lattice spacing along the $y$ direction, so that both $y_+$ and $y_-$ are equal to $y_x$, the initial $y$-coordinate of the $p_x$ monopole, up to multiples of $p$. As a result,
\begin{align}
\phi_x = \frac{2\pi q p_x}{p} (y_x - y_q) + 2\pi n    
\end{align}
where $n \in \mathbb{Z}$, which matches the phase derived from the lattice model (see Table \ref{tab:statist}). The same consideration also gives $\phi_y = (2\pi q p_y /p ) (x_q - x_y)$ mod $2\pi$. Cast as integrals of a gauge-invariant magnetic field, the two expressions in Eq. (\ref{eq:phixphiy}) are manifestly gauge-invariant. In the Appendix we show how to convert the area integrals in Eq. (\ref{eq:phixphiy}) to line integrals of some emergent vector potentials.

\section{Discussion}
\label{sec:10}

We have carefully revisited the Higgsing process proposed in Refs.~\cite{hermele18b, barkeshli18} to come up with an easy-to-implement recipe for constructing a stabilizer spin model from a given LGT. The Higgsing procedure is applied to the rank-2 U(1) LGT defined on a two-dimensional square lattice to obtain the rank-2 $\mathbb{Z}_p$ toric code. The monopole and dipole quatiparticles are  the elementary excitations of the R2TC. Allowed motions of both monopoles and dipoles in the R2TC model were analyzed in detail. 

The GSD of R2TC model varies from $p^3$ to $p^6$ depending on the commensurability of the lattice size with $p$, the local Hilbert space dimension. The counting of GSD was performed by examining the number of independent stabilizers as well as the number of independent logical operators. Our interpretation of the logical operators in terms of monopole-pair and dipole-pair creation and re-annihilation processes may shed further light on previous reports on the system size dependence of the GSD~\cite{hsieh17prb, barkeshli18, hermele18b}. 

The braiding statistics between a pair of monopoles and a monopole-dipole pair was worked out. We propose a new integral formula capturing the braiding statistics, which invites further field-theoretical investigation of the emergent quasiparticle dynamics. Checking the stability of our spin model remains as a future work, which can be done by perturbative continuous unitary transformation method previously exploited in the study of the phase diagram of the toric code or fracton models~\cite{vidal09prb,muhlhauser20prb}.

\acknowledgments H. J. H. was supported by the Quantum Computing Development Program (No. 2019M3E4A1080227). E.-G.M. was supported by National Research Foundation of Korea under Grant NRF-2019M3E4A1080411, NRF-2020R1A4A3079707, and NRF-2021R1A2C4001847.

\appendix*
\section{Aharonov-Bohm Phases for Monopole-monopole Braiding}
\label{sec:a}

The following integral expressions for the statistical phases were introduced in Sec. \ref{sec:9}:
\begin{align}
    \phi_x & = p_x \int_{x_-}^{x_+} dx \int_{y_-}^{y_+} dy \left[ \int_{y_0}^y B (x,y') dy'  \right] \nn
    \phi_y & = - p_y \int_{x_-}^{x_+} dx \int_{y_-}^{y_+} dy \left[ \int_{x_0}^x B (x' ,y) dx'  \right]. 
    \label{eq:AB-phase-B}
\end{align}
It is possible to apply Stokes' theorem and express the above surface integrals as line integrals of some effective vector potentials. First one writes the magnetic field $B(x,y)$ in the rank-2 U(1) gauge theory as the curl
\begin{align}
    B(x,y) = \partial_x C^y(x,y) - \partial_y C^x(x,y),
\end{align}
where ${\bm C}(x,y)=(C^x,C^y)$ is given by
\begin{align}
    C^x(x,y) & =(1 - \gamma) \partial_x A^{xy}(x,y) -\partial_y A^{xx}(x,y),\nn 
    C^y(x,y) & = \partial_x A^{yy}(x,y)-\gamma\partial_yA^{xy}(x,y)
\end{align}
for some arbitrary constant $\gamma$. Accordingly, one can write 
\begin{align}
\int^y_{y_0} B(x,y') dy' & = \partial_x \alpha_x^y -\partial_y \alpha_x^x , \nn 
\int^x_{x_0} B(x',y) dx' & = \partial_x \alpha_y^y -\partial_x \alpha_y^x 
\end{align}
where
\begin{align}
\alpha^x_x & = \int^y_{y_0} dy' \left[ C^x(x,y') - C^x(x,y_0)\right] , \nn 
\alpha^y_x & = \int^y_{y_0} dy' C^y(x,y')  \nn 
\alpha^x_y & = \int^x_{x_0} dx' C^x(x',y) \nn
\alpha^y_y & = \int^x_{x_0} dx' \left[ C^y(x',y) - C^y(x_0,y) \right] . 
\end{align}
%
Thus we obtain a pair of emergent vector potentials ${\bm \alpha}_x=(\alpha_x^x,\alpha_x^y)$ and ${\bm \alpha}_y=(\alpha_y^x,\alpha_y^y)$ with which to express the AB phases as line integrals 
\begin{align}
    \phi_x & = p_x \oint d {\bm r} \cdot {\bm \alpha}_x , ~~~ \phi_y = - p_y \oint d{\bm r} \cdot {\bm \alpha}_y .
    \label{eq:AB-phase-a}
\end{align}
The gauge invariance of $\phi_x , \phi_y$ under the transformation 
\begin{align}
    A^{ab}  & \rightarrow A^{ab} + \partial_a f^a &&(a = b)  \nn 
    & \rightarrow A^{ab} + \partial_a f^b + \partial_b f^a && (a \neq b) 
\end{align}
follows readily. 

%
%

One way to write down the tensor $A^{ab}$ for the localized magnetic field $B({\bm r}) =( 2\pi q /p )  \delta^2({\bm r} - {\bm r}_q )$ was suggested in Eq. (\ref{eq:Aab-for-localized-B}) in Sec. \ref{sec:9}. Alternative ways of writing the tensor potential for the same magnetic field are  
\begin{align}
& A'^{xx}({\bm r}) = A'^{yy}({\bm r}) = 0, \nn
& A'^{xy}({\bm r}) = -\frac{2 \pi q}{p} \theta(x-x_q) \theta(y-y_q),
\end{align}
and
\begin{align}
&A''^{xx}({\bm r}) = A''^{xy}({\bm r}) = 0 \nn
&A''^{yy}({\bm r}) = \frac{2 \pi q}{p}  \left(x - x_q \right) \theta(x-x_q) \delta(y-y_q).
\end{align}

\bibliography{SC}
\end{document}